\documentclass[nohyper]{JHEP3}
\usepackage{cite}
\usepackage{epsfig}

\newcommand{\mycomm}[1]{\hfill\break $\phantom{a}$\kern-3.5em{\tt===$>$ \bf #1}\hfill\break}
\newcommand{\mycommA}[1]{\hfill\break $\phantom{a}$\kern-3.5em{\tt   $>$ \bf #1}\hfill\break}

\newcommand{\be}{\begin{equation}}
\newcommand{\ee}{\end{equation}}
\newcommand{\ba}{\begin{eqnarray}}
\newcommand{\ea}{\end{eqnarray}}

\def\tHooft{\hbox{\tiny 't Hooft}}
\def\eq#1{Eq.~(\ref{#1})}

\def\MSbar{\hbox{\tiny ${\overline{\rm MS}}$}}

\def\PT{\hbox{\tiny PT}}

\catcode`\@=11 
\def\lsim{\mathrel{\mathpalette\@versim<}}
\def\gsim{\mathrel{\mathpalette\@versim>}}
\def\@versim#1#2{\vcenter{\offinterlineskip
        \ialign{$\m@th#1\hfil##\hfil$\crcr#2\crcr\sim\crcr } }}
\catcode`\@=12 

\newcommand{\nsl}{\mbox{$n$\hspace{-0.5em}\raisebox{0.1ex}{$/$}}}

\title{On the quark distribution in an on-shell heavy quark and its
all--order relations with the perturbative fragmentation function}

\author{Einan Gardi \\
Cavendish Laboratory, University of Cambridge\\
Madingley Road, Cambridge, CB3 0HE, UK}

\abstract{I present new results on the quark distribution in an
on-shell heavy quark in perturbative QCD and explore its all--order
relations with heavy--quark fragmentation. I~first compute the momentum
distribution function to all orders in
the large--$\beta_0$ limit and show that it is identical to the
perturbative heavy--quark fragmentation function in the same
 approximation. I then analyze the Sudakov limit of the distribution
and the fragmentation functions using Wilson lines
and prove that the corresponding Sudakov exponents in the non-Abelian theory
are the same to any logarithmic accuracy.
The anomalous dimension is then determined to two--loop order,
corresponding to next--to--next--to--leading logarithmic accuracy
in the exponent, in two ways: the first by extracting the singular
terms from a recent calculation of the fragmentation function
and the second by performing the two--loop Wilson--line calculation
in configuration space. I find perfect agreement between the two.}

\keywords{QCD, inclusive B decay, fragmentation, heavy quarks, renormalons, Sudakov resummation}

\preprint{Cavendish-HEP-05/03}

\begin{document}

\vskip 50pt
\section{Introduction}

Relations between space--like processes that probe the parton distribution
function and time--like processes involving fragmentation have been
known since long. This includes the Drell--Yan--Levy relation between
deep--inelastic structure functions and single--particle inclusive cross section in
$e^+e^-$ annihilation~\cite{DYL},
as well as the Gribov--Lipatov reciprocity relation~\cite{GL},
relating the space--like and time--like
splitting functions that determine the
Dokshitzer--Gribov--Lipatov--Altarelli--Parisi (DGLAP)~\cite{DGLAP}
evolution of parton densities and
fragmentation functions, respectively. In dimensional regularization,
the latter relation is violated beyond the leading order.

In this paper we find and explore new all--order perturbative relations between distribution and fragmentation
functions, which are specific to heavy quarks\footnote{We do not
consider in this paper deep inelastic structure functions.}.
When dealing with heavy quarks one can define and compute perturbative
distribution and fragmentation functions, replacing
the hadronic states by ones composed of on-shell quarks and gluons.
Owing to collinear singularities this is not possible for light quarks, where only
the ultraviolet (DGLAP) evolution of these functions can be computed
in perturbation theory, while the actual matrix elements, which set the
initial condition for the evolution, can only be defined non-perturbatively.
For heavy quarks the mass effectively cuts off
collinear radiation and, if
$m\gg \Lambda$, it provides a hard scale for the coupling. This makes the
perturbative initial condition for DGLAP evolution of the distribution and
fragmentation functions well defined to any order
in perturbation theory. These perturbative initial
conditions are the subject of this paper. Of course, these functions still differ
from their non-perturbative counterparts by power corrections, which we shall
briefly discuss as well.

The non-perturbative process--independent definitions of distribution and fragmentation
functions~\cite{CS,Collins} --- see Eqs.~(\ref{definition}) and~(\ref{D_def}),
respectively --- apply to both light and  heavy quarks.
These field--theoretic definitions are based on the Fourier transform of certain non-local
matrix elements on the lightcone.
The distribution function $f(x;\mu)$ measures the longitudinal momentum distribution
of a quark in a hadron, where the quark momentum fraction is $x\in [0,1]$. This
function can be interpreted as a probability distribution, barring the fact that it
requires renormalization.
Similarly, the fragmentation function $d(x;\mu)$ measures the probability
distribution of producing a hadron with fraction $x$ of the momentum originally
carried by the quark. Also here $x\in [0,1]$.
Using the heavy quark expansion~\cite{JR} one can show that for $\mu \sim m$ both
$f(x;\mu)$ and $d(x;\mu)$ peak at $1-x={\cal O} (\Lambda/m)$.

The momentum distribution function of a heavy quark in a heavy meson plays
a central role in the calculation of inclusive decay
spectra such as $\bar{B}\longrightarrow X_s\gamma$ and semileptonic
$B$ decays~\cite{Neubert:1993um,Bigi:1993ex,Falk:1993vb,KS,Bauer:2003pi,
Bosch:2004th,Neubert:2004dd,BDK}. This distribution essentially determines the shape of
inclusive decay spectra in the experimentally--important endpoint region.
In this kinematic domain the invariant mass of the produced hadronic system is
small compared to the quark mass and the ${\cal O}(\Lambda/M)$ fraction of
the momentum which is carried by the light degrees of freedom in the meson
becomes important.
In this context it is convenient to analyze the quark distribution
function in the infinite--mass limit, where it is often called the
``Shape Function''.

Although the quark distribution in a heavy meson is a non-perturbative
distribution, it has an important perturbative ingredient.
Being completely inclusive, this distribution can be approximated by the quark
distribution in an on-shell heavy quark $f_{\PT}(x;\mu)$,
while non-perturbative effects
 are treated as corrections. This idea is theoretically appealing,
 however, it is not easy to achieve.
First, the perturbative description of the quark
distribution in an on-shell heavy quark can only be reliable if
Sudakov logarithms are resummed, resummation that takes
the form of exponentiation in moment space.
Moreover, running--coupling effects are significant and lead to infrared renormalons
in these moments. The corresponding ambiguities are only
resolved at the non-perturbative level, making the resummation of
running--coupling effects necessary for systematic separation between
perturbative and non-perturbative contributions.
The Dressed Gluon {\mbox Exponentiation} (DGE)
approach~\cite{BDK,DGE_thrust,Gardi:2001di,CG,AG} incorporates
renormalon resummation in the calculation of the Sudakov exponent. This opens up the
way for consistently using the resummed quark distribution in an on-shell quark as
the baseline for describing the quark distribution in a heavy meson.

In a completely different application of QCD, namely heavy--quark production
in hard scattering processes, one encounters the
heavy--quark fragmentation function, $d(x;\mu)$.
Also here the resummed
perturbative initial condition for the evolution, $d_{\PT}(x;\mu)$, provides a
baseline for a
systematic description of the non-perturbative
object~\cite{MN,Webber_Nason,CC,CG,MM}.
DGE--based predictions~\cite{CG}, which involve just one or
two non-perturbative power corrections, have proven successful
in describing the inclusive $b$ production cross section measured at LEP.

In this paper we compute the quark distribution in an on-shell
heavy quark, $f_{\PT}(x;\mu)$, in two
different limits. In each case a certain gauge--invariant
 set of radiative corrections is
controlled to all orders. The two limits are:
\begin{itemize}
\item{} The single--dressed--gluon approximation (Sec.~\ref{SDG}). Taking
the large--$\beta_0$ limit --- or, formally, the large--$N_f$ limit --- we perform an
all--order resummation of running coupling effects (renormalons).
This resummation has two
applications~\cite{Gardi:1999dq,Dokshitzer:1995qm,Beneke:2000kc}:
first, it improves the leading--order
calculation incorporating BLM-type radiative corrections to
all orders~\cite{BLM,Brodsky:2000cr}, and second,
it probes the structure of power corrections.
\item{} The $x\longrightarrow 1$ Sudakov limit (Sec.~\ref{Sudakov_limit}).
Taking the large--$x$ limit we keep only singular terms in $f_{\PT}(x;\mu)$,
which build up the Sudakov exponent. Here the calculation is done to two-loop order,
corresponding to next--to--next--to--leading logarithmic (NNLL) accuracy in the
exponent.
More generally, following the work of Korchemsky and Marchesini in Ref.~\cite{KM},
we show that the singular $x\longrightarrow 1$ terms are fully captured by
the distribution defined in the $m\longrightarrow \infty$ limit, the so-called ``Shape Function''.
This means that the exponent is computable, to all orders, through the renormalization of a
$\Pi$-shaped Wilson--line operator\footnote{Recall that not all
Sudakov anomalous dimensions can be computed in the Eikonal approximation. For example,
the jet function controlling the large-$x$ limit of deep-inelastic
structure functions~\cite{Sterman:1986aj,CT,Gardi:2001di,GR}, which is sensitive to collinear
radiation from a light quark, cannot be reproduced in this approximation.} with two
antiparallel timelike rays,
connected by a lightlike segment.
\end{itemize}
Interestingly, in both limits we find that exactly the same resummed
expressions hold for the perturbative heavy--quark fragmentation function~\cite{CC,CG,MM}.
In general though, the two functions do differ; for example, their DGLAP evolution
away from the large--$x$ limit starts differing at two-loops~\cite{GL}.

\section{Quark distribution in an on-shell heavy quark computed with a
single dressed gluon\label{SDG}}

\subsection{Definition and calculation of the gluon emission cut}

We define the parton distribution~\cite{CS} in a heavy meson $\vert H(p)
\rangle$ in the standard way, by
\begin{equation}
\label{definition}
   f(x;\mu)=\int_{-\infty_{\,}}^{\infty}\frac{dy^{-}}{4\pi}\, {\rm
   e}^{-ixp^{+}y^{-}}\,
   \left< H(p) \right\vert\bar{\Psi}(y) \Phi_{y} (0,y) \gamma_{+} \Psi(0)
    \left\vert H(p)
    \right>_{\mu},
\end{equation}
where $\mu$ is introduced as an ultraviolet
renormalization scale for the operator, $y$ is a lightlike vector in the
``$-$'' direction, $\Phi_{y}
(0,y)$ is a path--ordered exponential in this direction connecting
the points $y$ and $0$, i.e.
\[
\Phi_{y}(0,y) \equiv {\bf P} \exp\left(ig\int_{0}^{y}dz_{\mu}
A_{\mu}(z)\right).\]  Some properties of $\Phi_{y}(0,y)$
are collected in Appendix~\ref{Wilson_lines}.

In order to compute the matrix element it is convenient use the
lightcone gauge, where
\begin{equation}\label{lightcone_gauge}
    \sum_{\lambda}
    \epsilon_{\mu}^{(\lambda)}\epsilon_{\nu}^{(\lambda)*}
    =-g_{\mu\nu}+\frac{k_{\mu}y_{\nu}+k_{\nu}y_{\mu}}{k\cdot y}.
\end{equation}
Then, at one loop there is only one diagram: the box. The
calculation, based on of \eq{definition} is summarized in Appendix
A. In $4-2\epsilon$ dimensions the result for the
not--yet--renormalized quark distribution is:
\begin{equation}\label{f_result}
    f^{(4-2\epsilon)}_{\PT}(x;\mu)=\delta(1-x)\left[ 1+{\cal O}(\alpha_s)\right] +
    \left(\frac{\mu^2{\rm e}^{\gamma_E}}{m^2}\right)^\epsilon\,
    \frac{C_F}{\beta_0}\,\int_0^{\infty} \,
du\,T(u)\,\left(\frac{\Lambda^2}{m^2}\right)^u\,B(x,u,\epsilon)
\end{equation}
with
\begin{equation}\label{borel}
    B(x,u,\epsilon)=\frac{\Gamma(\epsilon+u)}{\Gamma(1+u)}\,{\rm e}^{\frac53 u}\,
    (1-x)^{-2\epsilon-2u}x^u\,\left[ (1-u-\epsilon)\frac{x}{1-x}
    +\frac{1+u}{2}(1-x)\right],
\end{equation}
where we used the Borel--modified gluon propagator $1/(-k^2)
\longrightarrow 1/(-k^2)^{1+u}$ in order to resum running--coupling
effects of all orders. With one-loop running coupling $T(u)=1$. To
go beyond one--loop running we use the scheme--invariant Borel
representation~\cite{Grunberg:1992hf} where $T(u)$ is the Laplace transform
of the 't~Hooft
coupling:
\begin{eqnarray}
\label{tHooft_coupling}
A(\mu)&=&\frac{\beta_0\alpha_s^{\tHooft}(\mu)}{\pi}=
\int_0^{\infty}{du} \,T(u)\, \left(\frac{\Lambda^2}{\mu^2}\right)^u;
\qquad \qquad \frac{dA}{d\ln \mu^2} =-A^2(1+\delta A),
\nonumber \\
T(u)&=&\frac{(u\delta)^{u\delta}{\rm
e}^{-u\delta}}{\Gamma(1+u\delta)};\qquad \qquad \ln
(\mu^2/\Lambda^2)=\frac{1}{A}-\delta\ln\left(1+\frac{1}{\delta
A}\right)
\end{eqnarray}
with $\delta\equiv \beta_1/\beta_0^2$, where $\beta_0=
\frac{11}{12}C_A-\frac{1}{6}N_f$ and $\beta_1 =
\frac{17}{24}C_A^2-\frac{1}{8}C_FN_f-\frac{5}{24}C_AN_f$. We define
$\Lambda$ in ${\overline{\rm MS}}$.

Out of the terms in the square brackets in \eq{borel} only
$\frac{x}{1-x}$ (with a coefficient 1) comes from the Axial gauge
part of the propagator in \eq{lightcone_gauge}, which is
proportional to $1/k \cdot y$. Through the $y^{-}$ and the momentum
integration this singularity of the propagator is
converted into a $1/(1-x)$ singularity,
$k\cdot y/p \cdot y\longrightarrow (1-x)$. The square brackets are a
generalization of the splitting function. It is noted that
\eq{f_result} is {\em identical} to the {\em perturbative
fragmentation function} \cite{CC} computed in the same
approximation in Ref.~\cite{CG} using a different technique.

\subsection{Moments and evolution}

In \eq{f_result} we computed only the real gluon emission
contribution. Clearly this result is singular at $x\longrightarrow
1$. Virtual corrections generate ${\cal O}(\alpha_s)$ divergent
contributions, which are proportional to $\delta(1-x)$. Owing  to
the inclusive nature of \eq{definition} infrared singularities
cancel out. However, there are also ultraviolet singularities. We
therefore proceed by first taking a logarithmic derivative with
respect to the mass to remove the ultraviolet divergence. Then we
can go to four dimensions setting $\epsilon=0$. Next, we go to
moment space,
\begin{equation}
\label{F_N}
F_N=\int_0^1dx x^{N-1} f(x;\mu),
\end{equation}
where we can use the fact that the first moment $F_{N=1}$
corresponds to the total number of quarks in the quark, a conserved
current, so it is $F_{N=1}=1$ for any mass and the derivative must
vanish identically. This allows us to reconstruct the virtual
($N$--independent) terms missing in \eq{f_result} out of the
real--emission ($N$--dependent) ones. We get:
\begin{eqnarray}
\label{N_ren_sum} &&\hspace*{-30pt} \frac{d\ln F_{N}}{d\ln m^2} =
-\frac{ C_F}{\beta_0}\,\int_0^{\infty}du\,
\,T(u)\,\left(\frac{\Lambda^2}{m^2}\right)^u\,B_{F}(N,u),
\end{eqnarray}
with
\begin{eqnarray}
\label{B_F} B_{F}(N,u)&=&\,
\int_{0}^1\,dx\,\left(x^{N-1}-1\right)\,B(x,u,\epsilon=0)=\nonumber \\
&&\hspace*{-40pt} {\rm e}^{\frac53 u} \,\left[ (1-u)\Gamma(-2u)
\left(\frac{\Gamma(N+1+u)}{\Gamma(N+1-u)}-\frac{\Gamma(2+u)}{\Gamma(2-u)}
\right)\right. \\ \nonumber && \hspace*{50pt}\left.+
\frac12(1+u)\Gamma(2-2u)\left(\frac{\Gamma(N+u)}{\Gamma(N+2-u)}-
\frac{\Gamma(1+u)}{\Gamma(3-u)}\right)\right]+\,{\cal O}(u/\beta_0).
\end{eqnarray}

Next, we would like to integrate over $\ln m^2$ to recover $F_N$.
However, this brings back the ultraviolet divergence which takes the
form of a $1/u$ singularity. Therefore, we must perform ultraviolet
subtraction. The result is:
\begin{eqnarray}
\label{N_space_result}
 F_N^{\PT}(m;\mu_{F})\,=\, 1\,&+&\,\frac{ C_F}{\beta_0}\,\int_0^{\infty}\,\frac{du}{u} \,T(u)\,
\left(\frac{\Lambda^2}{m^2}\right)^{u}\,\left[B_{F}(N,u)\,+\,
\left(\frac{m^2}{\mu_{F}^{2}}\right)^{u}\,B_{\rm AP}(N,u) \right]\nonumber\\
\hspace*{0pt}&+&\,{\cal O}(1/\beta_0^2),
\end{eqnarray}
where~$\mu_F$ is a factorization scale and~$B_{\rm AP}(N,u)$ is the
(non-singlet) Altarelli-Parisi evolution kernel, \be \label{B_AP}
B_{\rm
AP}(N,u)\,=\,\sum_{n=0}^{\infty}\frac{\gamma_n(N)\,u^{n}}{n!}. \ee
The leading order coefficient in~(\ref{B_AP}) is
renormalization--scheme invariant, and it equals to the~$u=0$ limit
of~(\ref{B_F}), ensuring the cancellation of the $1/u$ singularity
in~(\ref{N_space_result}):
\[
\gamma_0(N)=S_1(N)-\frac{3}{4}+\frac12\left(\frac{1}{N+1}-\frac{1}{N}\right),
\]
where $S_k(N)\equiv \sum_{j=1}^N {1}/{j^k}$, so
$S_1(N)=\Psi(N+1)+\gamma_E$. In the $\overline{\rm MS}$
factorization scheme~$\gamma_n(N)$ are known to NNLO ($n=2$) in
full~\cite{MVV} and to all orders in the large-$\beta_0$
limit~\cite{Gracey}, see Appendix \ref{sec:cusp}.

We find that the final result for the moments of the quark
distribution in the large--$\beta_0$ limit, given by
\eq{N_space_result}, is identical to the one corresponding to the
heavy-quark fragmentation in the same approximation, see Eqs. (38)
and (39) in Ref.~\cite{CG}. The expansion of $F_N^{\PT}(m;\mu_{F})$
in powers of $\alpha_s$ can be readily obtained from \eq{N_space_result} by expanding
the terms in the square brackets in powers of $u$ and integrating order by order. The result,
to ${\cal O}(\alpha_s^2)$, appears in Eq. (43) of Ref.~\cite{CG}.

The infrared--renormalon structure can be read off \eq{B_F}: there are simple
poles at all integer and half integer values of $u$, except for $u=1$. Note that all
these singularities are absent in the real--emission expression in $x$ space, where
one finds just an upper limit of the form $(1-x)\gsim \Lambda/m$
on the values of $x$ for which the Borel integral exists. These
renormalons are all associated with the integration over $x$ for
$x\longrightarrow 1$, and within the scheme--invariant
formulation of the Borel transform (where $\beta_1$--terms are factored out)
they are expected to remain simple
poles in the full theory. As already noted in Refs.~\cite{CG,BDK},
the residues in \eq{B_F}
depend on~$N$ such that at large~$N$ ambiguities appear as integer powers
of $(N\Lambda/m)$. These parametrically--enhanced power terms are related to
renormalon ambiguities in the Sudakov exponent, and they will be discussed further
in the next section.

\section{Sudakov logs in the quark distribution function\label{Sudakov_limit}}

\subsection{Definitions and results\label{sec:results}}

In the large--$N$ limit Sudakov logarithms exponentiate as follows:
\be
F_N^{\PT}(m;\mu_{F})=\tilde{H}(m;\mu_F)\,\tilde{S}_N(m;\mu_F)\,+\,{\cal
O}(1/N), \label{soft_mom_def} \ee where
\begin{eqnarray}
\label{Soft_Catani} \hspace*{-20pt}\tilde{S}_N(m;\mu_F) &=&
\exp\Bigg\{\sum_{n=1}^\infty \left(\frac{\alpha_s(m)}{\pi}\right)^n
\,\,\,\left[\sum_{k=1}^{n+1} C_{n,k} \ln^kN +{\cal O}(1) \right]\Bigg\}\\
\nonumber &=&\exp \Bigg\{ \int_0^1 dx \,\frac{x^{N-1}-1}{1-x}\,
\left[\int_{(1-x) m}^{\mu_F}\frac{2d\mu}{\mu}{\cal
A}\big(\alpha_s(\mu)\big)\,-\,{\cal D}\big(\alpha_s((1-x)
m)\big)\right]\Bigg\}.\\\nonumber
\end{eqnarray}
Here\footnote{Notations in the literature vary; for example in Ref.
\cite{CC}, Eq. (69), our ${\cal A}$ and ${\cal D}$ are denoted by
$A$ and~$-H$, respectively.} ${\cal A}$ and ${\cal D}$ are Sudakov
anomalous dimensions having the following perturbative expansions in
the ${\overline{\rm MS}}$ renormalization scheme:
\begin{eqnarray}
\label{A_cusp_expansion} {\cal A}\big(\alpha_s(\mu)\big)&=&
\sum_{n=1}^{\infty}
A_n^{\MSbar}\left(\frac{\alpha_s^{\MSbar}(\mu)}{\pi}\right)^n\,=\,
\frac{C_F}{\beta_0}\sum_{n=1}^{\infty} a_n^{\MSbar}
\left(\frac{\beta_0\alpha_s^{\MSbar}(\mu)}{\pi}\right)^n,\nonumber
\\
{\cal D}\big(\alpha_s(\mu)\big)&=&
\sum_{n=1}^{\infty}
D_n^{\MSbar}\left(\frac{\alpha_s^{\MSbar}(\mu)}{\pi}\right)^n\,=\,
\frac{C_F}{\beta_0}\sum_{n=1}^{\infty} d_n^{\MSbar}
\left(\frac{\beta_0\alpha_s^{\MSbar}(\mu)}{\pi}\right)^n,
\end{eqnarray}
where the normalization\footnote{We shall mostly use the notation
$a_n$ and $d_n$ for the coefficients. This is convenient for comparison with the
large--$\beta_0$ limit and with the Borel formulation but it does not imply any
additional approximation: $a_n$ and $d_n$ contain all color factors.}
is such that $a_1=d_1=1$. The coefficients $a_2^{\MSbar}$ and
$a_3^{\MSbar}$ are known based on calculations in deep inelastic
scattering (see e.g.~\cite{MVV}). They are given in \eq{a23_MSbar}.
$d_2^{\MSbar}$ is given by:
\begin{equation}
d_2^{\MSbar}= \frac {1}{9}  + {\displaystyle \frac{C_A}{{\beta
_{0}}} {\left({\displaystyle \frac {9}{4}} \,\zeta_3
 - {\displaystyle \frac {\pi ^{2}}{12}}  - {\displaystyle \frac {
11}{18}} \right)\,}}. \label{d_2}
\end{equation}
We shall explain how it was determined in Sec. \ref{sec:KM}.

${\cal A}\big(\alpha_s(\mu)\big)$ is the universal cusp anomalous
dimension which generates the double logs. It is well known that
this object can be defined through the renormalization of a Wilson
line with a cusp \cite{KR87,Korchemsky:1988si,Korchemskaya:1992je,KR92,KM,Grozin:1994ni,Beneke:1995pq}.
${\cal D}\big(\alpha_s(\mu)\big)$ is another anomalous dimension, which
describes soft radiation from the heavy quark and it is specific to
the heavy--quark distribution function.
In Sec.~\ref{sec:relation_to_frag} we shall prove that
the same anomalous dimension controls the large--$N$ limit of the
heavy--quark fragmentation function.
${\cal D}\big(\alpha_s(\mu)\big)$ too can be defined and computed using a
Wilson--line operator. In contrast with the cusp anomalous
dimension, it is associated with a {\em specific} $\Pi$--shaped
configuration with a finite light--like section to which two
antiparallel infinite time--like rays attach. This object was first
analyzed by Korchemsky and Marchesini in Ref.~\cite{KM}.

The single--dressed--gluon calculation of the previous section has
just leading logarithmic accuracy: \eq{B_F} (or \eq{B_DJ_large_beta0} below)
fixes $C_{n,k=n+1}$ for any $n$, but gives only the part of the
$C_{n,k\leq n}$ which is leading in $\beta_0$. In the full theory
there are additional terms in the exponent which have different
color factors. While ${\cal A}$ (the cusp anomalous dimension)
contains both non-Abelian $C_A/\beta_0$ and Abelian $C_F/\beta_0$
terms, ${\cal D}$ has only non-Abelian ones. These coefficients can
be determined by computing the anomalous dimensions order by order.
Next--to--leading logarithmic accuracy ($C_{n,k\geq n}$) requires
fixing $a_2$ and $d_1$. The state of the art is
next--to--next--to--leading logarithmic accuracy ($C_{n,k\geq n-1}$)
which requires knowing also~$a_3$ and~$d_2$. Note that
the calculation of the exponent to this formal
accuracy should be complemented by full next--to--next--to--leading
order calculations of the hard coefficient function. These are yet
unavailable for radiative or semileptonic decay spectra.

Switching to the scheme invariant Borel representation of the two
anomalous dimensions,
\begin{eqnarray}
{\cal A}\big(\alpha_s(\mu)\big)= \frac{C_F}{\beta_0}\int_0^{\infty}
du \,T(u)
\,\left(\frac{\Lambda^2}{\mu^2}\right)^u\, B_{\cal A}(u),\nonumber \\
{\cal D}\big(\alpha_s(\mu)\big)=  \frac{C_F}{\beta_0}\int_0^{\infty}
du \,T(u) \,\left(\frac{\Lambda^2}{\mu^2}\right)^u\, B_{\cal D}(u),
\end{eqnarray}
the moments of the distribution function become:
 \be F_N^{\PT}(m;\mu_{F})=H(m;\mu_F)\,S_N(m;\mu_F)\,+\,{\cal
O}(1/N), \label{soft_mom_def_my} \ee
with\footnote{Although both
\eq{Soft_Catani} and \eq{Soft} are normalized such that the $N=1$
moment is identically unity, the integration over $x$ in
\eq{Soft_Catani} generates, in addition to the relevant $\ln N$
terms, some finite terms along with terms that vanish as powers of
$1/N$. For this reason the functions in \eq{soft_mom_def} have a
tilde distinguishing them from those of \eq{soft_mom_def_my}. As far
as the logarithms are concerned the exponent in \eq{Soft} is
identical to that of \eq{Soft_Catani}.}
\begin{eqnarray}
\label{Soft} &&\hspace*{-10pt}S_N(m;\mu_F)=
\exp\bigg\{\sum_{n=1}^\infty \left(\frac{\alpha_s(m)}{\pi}\right)^n
\,\,\,\sum_{k=1}^{n+1}
C_{n,k} \ln^kN \bigg\}=\\
\nonumber &&\exp \bigg\{
\frac{C_F}{\beta_0}\int_0^{\infty}\frac{du}{u} \,T(u)\,
\left(\frac{\Lambda^2}{m^2}\right)^u\,\,
\bigg[ B_{\cal S}(u)\Gamma(-2u)\left(N^{2u}-1\right)
+\left(\frac{m^2}{\mu_{F}^2}\right)^u B_{\cal A}(u)\ln N\bigg]
\bigg\},
\end{eqnarray}
where $C_{n,k=1}$ depend on $\ln \mu_F/m$ while $C_{n,k>1}$ are
numbers, and where
\begin{eqnarray}
\label{D_to_S} B_{\cal S}(u)&\equiv & B_{\cal A}(u) - u B_{\cal D}(u).
\end{eqnarray}
These expressions are completely general.

Based on \eq{B_F} we find that the hard function that incorporates the
finite terms at $N\longrightarrow \infty$
is given by:
\begin{eqnarray}
H(m;\mu_F)=1+\frac{C_F\alpha_s}{\pi}\left[\left(-\frac34+\gamma_{E}\right)\ln
\frac{m^{2}}{\mu_{F}^{2}}
+1-\frac{\pi^{2}}{6}+\gamma_E-\gamma_{E}^2\right] +{\cal
O}(\alpha_s^2), \label{H_}
\end{eqnarray}
and that in the large--$\beta_0$ limit the Borel transform of the
anomalous dimension $B_{\cal S}(u)$  is given by
\begin{eqnarray}
\label{B_DJ_large_beta0} B_{\cal S}(u)&=&{\rm e}^{\frac53
u}(1-u)\,+\,{\cal O}(1/\beta_0 ).
\end{eqnarray}
The information contained in Eqs.~(\ref{B_DJ_large_beta0}),
(\ref{D_to_S}) and (\ref{G_large_Nf}) can readily be translated into
values of the coefficients in \eq{A_cusp_expansion} in the
large--$\beta_0$ limit. One gets:
\begin{equation}
\label{dn_an_relation}
\left.d_n^{\MSbar}\right\vert_{{\rm large }\,
\beta_0}=\frac{1}{n}\bigg[\left. a_{n+1}^{\MSbar}\right\vert_{{\rm
large}\, \beta_0}
-\left(\frac53\right)^n\left(1-\frac{3n}{5}\right)\bigg].
\end{equation}
The first few orders are summarized in Table \ref{Table_coeff}. Note
that the convergence of the cusp anomalous dimension ${\cal A}$ is
much faster\footnote{In fact it is probably faster than all other
Sudakov anomalous dimension. See e.g. table 1 in \cite{GR}.} than
that of ${\cal D}$. This property presumably persists in the full
theory.
\begin{table}
\centering
\begin{tabular}{|l|l|l|l|}
\hline
 $n$& $\left. a_n^{\MSbar}\right\vert_{{\rm large}\, \beta_0}$
&
$\left. d_n^{\MSbar}\right\vert_{{\rm large}\, \beta_0}$& ratio\\
\hline 1& 1& 1& 1. \\ 
2& $\frac{5}{3}$& $\frac{1}{9}$& 15. \\ 
3& $- \frac{1}{3}$& $\frac{109}{81} - \frac{2}{3} \zeta_3$&
$-$0.6126 \\ 
4& $\frac{1}{3} -$2 $\zeta_3$& $\frac{1}{120} \pi^4 + \frac{212}{81}
- \frac{5}{6}$
 $\zeta_3$& $-$0.8531 \\
5& $\frac{1}{30} \pi^4 - \frac{1}{3} - \frac{10}{3} \zeta_3$&
$\frac{6331}{1215} - \frac{6}{5} \zeta_5 + \frac{1}{90} \pi^4 +
\frac{2}{15} \zeta_3$&
$-$0.2099 \\
6& $\frac{1}{3} -$6 $\zeta_5 + \frac{1}{18} \pi^4 + \frac{2}{3}
\zeta_3$& $- \frac{5}{3} \zeta_5 - \frac{1}{540} \pi^4 +
\frac{1}{567} \pi^6 + \frac{20191}{2187} + \frac{1}{3} \zeta_3^2 -
\frac{1}{9} \zeta_3$&
0.0347 \\
\hline
\end{tabular}
\caption{The large--$\beta_0$ part of the coefficients
$a_n^{\MSbar}$ and $d_n^{\MSbar}$ in \eq{A_cusp_expansion} and their
ratios.} \label{Table_coeff}
\end{table}

Expanding the Borel integral and the exponential in \eq{Soft} one
obtains the log-enhanced terms, order by order in the coupling. In
contrast with \eq{Soft_Catani}, \eq{Soft} incorporates renormalon
resummation in the Sudakov exponent (DGE). It exposes the fact that
the exponent contains infrared renormalon ambiguities: poles in
$\Gamma(-2u)$ along the integration path, which generate power--like
ambiguities~$\sim\,\left(N\Lambda/m\right)^j$, where $j$ are
integers. These parametrically--large ambiguities are indicative of
corresponding non-perturbative power terms.
As shown in Ref.~\cite{BDK} (see also \cite{Grozin:1994ni}),
the $u=\frac12$ renormalon is related to the leading infrared
renormalon in the pole mass~\cite{Beneke:1994sw,Bigi:1994em},
while higher powers are associated with other local
matrix elements which constitute the non-perturbative quark distribution
function~\cite{Bigi:1993ex} or ``Shape Function''~\cite{Neubert:1993um}.

In the DGE approach the Borel integration is performed
using the Cauchy Principal Value prescription, making an explicit power-like
separation between perturbative and non-perturbative contributions
to the exponent. Any information on the anomalous dimensions from
fixed--order calculations can be used in \eq{Soft}. Based on
Eqs.~(\ref{a23_MSbar}) and~(\ref{d_2}), we have:
\begin{eqnarray}
\label{BA_expanded} B_{\cal A}(u)&=&1+
\left(\frac53+c_2\right)\frac{u}{1!}+\left(-\frac 13+c_3\right)
\frac{u^2}{2!}+\,{\cal O}(u^3),\\
\label{B_DB} B_{\cal D}(u) &=& 1 +  \left[  \! {\displaystyle \frac
{1}{9}}  + {\displaystyle \frac{C_A}{{\beta _{0}}}
{\left({\displaystyle \frac {9}{4}} \,\zeta_3
 - {\displaystyle \frac {\pi ^{2}}{12}}  - {\displaystyle \frac {
11}{18}} \right)\,}}  \!  \right] \frac{u}{1!}\,+\,{\cal O}(u^2),
\end{eqnarray}
where $c_n$ represent contributions that are subleading in
$\beta_0$; $c_2$ and $c_3$ are given in Appendix~\ref{sec:cusp}.
Note that the relations of $c_n$ for $n\geq 3$ with $a_n^{\MSbar}$
involves   the coefficients of the $\beta$ function. The same is
true for the  ${\cal O}(u^2)$ and higher order coefficients in
$B_{\cal D}(u)$. Finally, note that since the analytic dependence on
$u$ is not known beyond the large--$\beta_0$ limit, there remains
some uncertainty in evaluating the $u$--integral in \eq{Soft} with
Eqs.~(\ref{BA_expanded}) and~(\ref{B_DB}). This issue is addressed
in detail in a forthcoming publication~\cite{AG}.

\subsection{Strictly factorized form and Wilson lines\label{sec:Wilson_lines}}

The factorization--scale dependence of the Sudakov factor of \eq{Soft}
(or \eq{Soft_Catani}) is governed by the cusp anomalous dimension alone
while its dependence on the soft scale $m/N$ is different.
Furthermore, by convention this factor is normalized such that its first
moment ($N=1$) is unity for any $\mu_F$; this
normalization is natural since the full quark distribution function indeed
obeys $F_{N=1}\equiv 1$.
These constraints can be satisfied only if the
exponent acquires some dependence on the hard scale $m$.
However, such dependence cannot be consistent with the Wilson--line operator
definition~\cite{KR87,KR92,KS,KM}, nor with the effective field
theory approach~\cite{Bauer:2000yr,Beneke:2002ph,Bauer:2003pi,Bosch:2004th,Neubert:2004dd}.
In the latter definitions the Sudakov factor cannot depend on $m$,
but only on the soft scale $m/N$ and on the factorization scale, the
renormalization scale of the operator. In order to convert \eq{Soft}
to a strictly factorized form we reshuffle radiative corrections
that depend only on the hard scales into the hard function, writing
\begin{eqnarray}
\hspace*{-30pt}
F_N^{\PT}(m;\mu_F)&=&H(m;\mu_F)\,S_N(m;\mu_F)+\,{\cal O}(1/N )
=
{\cal H}(m;\mu_F) \,{\cal S}(N \mu_F /m) +\,{\cal O}(1/N
)\label{strict_fact}
\end{eqnarray}
where the first expression represents the definition of
Sec.~\ref{sec:results} whereas the second assumes strict
factorization, as in Ref.~\cite{KM}. Although we do not write it
explicitly, both ${\cal H}$ and ${\cal S}$ depends also on
$\alpha_s(\mu)$, where the renormalization scale of the coupling,
$\mu$, is not necessarily equal to the factorization scale $\mu_F$.
In practice we shall be using the standard dimensional
regularization with $\mu$ defined in the ${\overline {\rm MS}}$
scheme for both the operator and the coupling.

We can now write the Borel representation of ${\cal S}(N\mu/m)$ as
follows:
\begin{eqnarray}
\label{cal_Soft} &&{\cal S}(N\mu/m)=  \exp\bigg\{\sum_{n=1}^\infty
\left(\frac{\alpha_s(\mu)}{\pi}\right)^n \,\,\,\sum_{k=1}^{n+1}
{\cal C}_{n,k} \ln^k (N\mu/m) \bigg\}=\\ \nonumber&&\exp \bigg\{
\frac{C_F}{\beta_0}\int_0^{\infty}\frac{du}{u} \,T(u)\,
\left(\frac{\Lambda^2}{\mu^2}\right)^u\,\,
\bigg[ B_{\cal S}(u)\Gamma(-2u)\left((N\mu/m)^{2u}-1\right)+ B_{\cal
A}(u)\ln (N\mu/m)\bigg] \bigg\}.
\end{eqnarray}
$B_{\cal S}(u)$ and $B_{\cal A}(u)$ are the same Borel functions
appearing in \eq{Soft}; clearly \eq{cal_Soft} differs from \eq{Soft}
just by $N$--independent terms, and it is therefore consistent with
\eq{strict_fact}.  Note that in contrast with~\eq{Soft}
the normalization (the $N=1$ moment) of the strictly--factorized Sudakov factor of
\eq{cal_Soft} strongly depends on $\ln\mu/m$ to any order in
perturbation theory.

Since there are no collinear singularities from the heavy--quark line,
Sudakov logarithms in ${\cal S}(N\mu/m)$ are related to soft gluons
and to gluons that are collinear to the lightcone direction $y$ in
\eq{definition}. One therefore expects that the log-enhanced terms
of the quark distribution function \eq{definition} would not change
if the heavy--quark lines are replaced by time--like Wilson lines.
The Wilson line describes the Eikonal interaction of the heavy quark
with soft gluons. This is equivalent to taking the quark mass to
infinity.

In order to convert \eq{definition} to a Wilson--line operator
definition \cite{KM} one applies the Eikonal approximation replacing
the interacting quark field $\Psi(z)$ by a free heavy quark field
$\psi(z)$ multiplied by a Wilson line in the direction $p$ which
extends to infinity, i.e.
\begin{equation}
\Psi(z)\longrightarrow \Phi_{p}(\infty,z) \psi(z)\,; \qquad
\bar{\Psi}(z)\longrightarrow \bar{\psi}(z)\Phi_{-p}(z,\infty)
\label{Quark_to_Wilson_line}
\end{equation}
with $\Phi_p(z_1,z_2)$ defined as in Appendix~\ref{Wilson_lines}.
Next, one uses the fact that the free fields annihilate the external
quark states to convert the matrix element to one in the vacuum.
Factoring out the Dirac structure one obtains:
\begin{eqnarray}
\label{Eikonalization} \hspace*{-25pt} \left< h(p)
\right\vert\bar{\Psi}(y) \Phi_{y} (0,y) \gamma_{+} \Psi(0)
\left\vert h(p)
    \right>_{\mu}\!\!\! \longrightarrow\, 2 p^{+}{\rm e}^{iy^{-}p^{+}}
\left< 0\right\vert {\Phi}_{-p}(y,\infty) \Phi_{y}
(0,y)\Phi_{p}(\infty,0)\left\vert 0 \right>_{\mu}.
\end{eqnarray}
The $x\longrightarrow 1$ singular terms in the quark distribution
function are therefore:
\begin{equation}
\label{definition_singular}
   \left. f_{\PT}(x;\mu)\right\vert_{x\longrightarrow 1}\sim
   \int_{-\infty}^{\infty}\frac{p^{+}dy^{-}}{2\pi}\, {\rm
   e}^{ip^{+}y^{-}(1-x)}\,
   \left< 0\right\vert {\Phi}_{-p}(y,\infty) \Phi_{y}
   (0,y)\Phi_{p}(\infty,0)\left\vert 0 \right>_{\mu},
\end{equation}

\begin{figure}[th]
\begin{center}
\epsfig{file=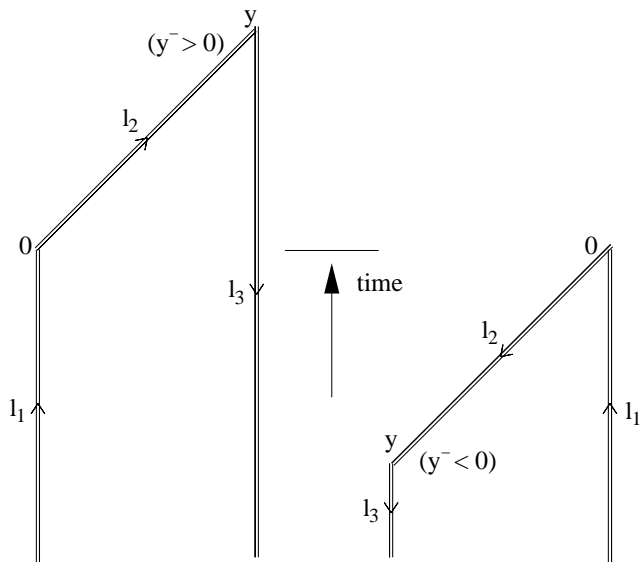,angle=0,width=8.4cm}
\caption{\label{fig:dist}Minkowski space-time picture (vertical axis as time and
horizontal axis as $x_3$) of the Wilson--line configuration
$W[C_S](ip\cdot y\mu/m)$ of~\eq{def_Wilson} representing
the quark distribution function in an on-shell quark in the infinite mass limit
(in the rest frame of this quark).
The two figures describe the situation
when $y^-$ is positive (l.h.s) or negative (r.h.s), where path--ordering on the
lightlike segment $l_2$ from $0$ to $y$ corresponds to
time--ordering and anti-time--ordering, respectively.}
\end{center}
\end{figure}

Following Ref.~\cite{KM} we define $W[C_S]$ by
\begin{eqnarray}
&&W[C_S](ip\cdot y  \mu /m,\alpha_s(\mu)) \equiv \left< 0\right\vert
{\Phi}_{-p}(y,\infty) \Phi_{y} (0,y)\Phi_{p}(\infty,0)\left\vert 0
\right>_{\mu};
 \label{def_Wilson}
\end{eqnarray}
it is shown in Fig.~\ref{fig:dist}.
$W[C_S](ip\cdot y  \mu /m,\alpha_s(\mu))$ is an analytic function
except on the negative real axis of
$ip\cdot y\mu/m$.
We note that, since $f_{\PT}(x;\mu)$ in \eq{definition_singular}
 is real, sign inversion of $p\cdot y$ is equivalent to complex
conjugation, i.e.
\begin{equation}
\label{complex_conjugation} W[C_S](ip\cdot y   \mu
/m,\alpha_s(\mu))=W[C_S]^{*}(-ip \cdot y  \mu /m,\alpha_s(\mu)).
\end{equation}

Up to ${\cal O}(1/N)$ corrections the Fourier integral in
\eq{definition} and the Mellin integral (\eq{F_N}) amount to
replacing $ip^{+}y^{-}$ by $N$~\cite{KM,Grozin:1994ni}. This can be
shown by first extending the $x$ integral to $-\infty$ relying on
the fact that the contribution is only from the $x$ near 1 region,
then performing the $x$ integral getting $-1/[ip^+y^--(N-1)]$, and
finally evaluating the $y^{-}$ integral by residue, closing the
contour through the lower half $p\cdot y$ plane. The result is
\begin{eqnarray}
\label{F_W_relation} F_N^{\PT}(m;\mu) =H_W^F(m;\mu) \,\times\, W[C_S](N  \mu
/m,\alpha_s(\mu))+{\cal O}(1/N),
\end{eqnarray}
where $H_W^F(m;\mu)$ is a hard coefficient function. It accounts for
finite terms at $N\longrightarrow \infty$ that are not captured by
the Eikonal approximation of \eq{Eikonalization}. The NLO expression
for this function is given in \eq{HW} below. $W[C_S]$ is therefore
related to ${\cal S}$ of \eq{cal_Soft} by
\begin{eqnarray}
&& {\cal S}(N  \mu /m)\equiv \frac{W[C_S](N \mu
/m,\alpha_s(\mu))}{W[C_S](1,\alpha_s(\mu))},
 \label{def_S_Wilson}
 \end{eqnarray}
where the additional normalization is needed to remove all
$N$-independent pieces, which are absent in \eq{cal_Soft} by
construction.

The Wilson line operator $W[C_S]$ was analyzed in detail in
Ref.~\cite{KM}, where its evolution equation was derived and a
two--loop calculation of the corresponding anomalous dimensions was
performed. Here we repeated this calculation finding:
\begin{eqnarray}
\label{MM_vs_KM} \ln W[C_S](N\mu/m,\alpha_s(\mu))&=&
 C_F \Big[- L^2 + L -\frac{5}{24}\pi^2\Big] \,\frac{\alpha_s(\mu)}{\pi}  \\
 &&\hspace*{-95pt} +\, C_F \Bigg[\left( - {\displaystyle
\frac {11\,\mathit{C_A}}{18}}  + {\displaystyle \frac {\mathit{N_f}
}{9}} \right)\,L^3 + \left(\left( - {\displaystyle \frac {17}{18} }
+ {\displaystyle \frac {\pi ^{2}}{12}} \right)\,\mathit{C_A} +
{\displaystyle \frac {\mathit{N_f}}{9}} \right)\,L^2 \nonumber \\
\nonumber
 &&\hspace*{-95pt}+\, \left(\left( - {\displaystyle \frac {55}{108}}  + {\displaystyle
\frac {9}{4}} \,\zeta_3 - {\displaystyle \frac {7\,\pi ^{2}}{18 }}
\right)\,\mathit{C_A} + \left( - {\displaystyle \frac {1}{54}}  +
{\displaystyle \frac {\pi ^{2}}{18}} \right)\,\mathit{N_f}\right)\,L
+{\cal O}(1) \Bigg] \left(\frac{\alpha_s(\mu)}{\pi}\right)^2+\cdots,
\end{eqnarray}
where $L=\ln N\tilde{\mu}/m$ with $\tilde{\mu}=\mu\,{\rm
e}^{\gamma_E}$ where $\mu$ is defined in the ${\overline {\rm MS}}$
scheme.

The {\em same} logarithmic terms can be obtained from \eq{cal_Soft}
by expressing it in terms of~$L$,
\[
\ln {\cal S}(N\mu/m)= 
\frac{C_F}{\beta_0}\int_0^{\infty}\frac{du}{u} \,T(u)\,
\left(\frac{\Lambda^2}{\mu^2}\right)^u\,\,
\hspace*{-5pt}\bigg[ B_{\cal S}(u)\Gamma(-2u)\left({\rm
e}^{2(L-\gamma_E)u}-1\right) + B_{\cal A}(u)(L-\gamma_E)\bigg] ,
\]
substituting the anomalous dimension coefficients of
Eqs.~(\ref{D_to_S}), (\ref{BA_expanded}) and (\ref{B_DB}) and
expanding. The $N$--independent terms, such as
$-\frac{5}{24}\pi^2C_F{\alpha_s(\mu)}/{\pi}$, are of course
different, see \eq{def_S_Wilson}. For $H_W^F(m;\mu)$ in
\eq{F_W_relation} we find:
\begin{equation}
\label{HW}
H_W^F(m;\mu)=1+C_F\frac{\alpha_s(\mu)}{\pi}\left[\ln^2\frac{m}{\mu}
-\frac12\ln\frac{m}{\mu}
+\frac{\pi^2}{24}+1\right] +\cdots.
\end{equation}

Our result in \eq{MM_vs_KM} agrees with that of Ref.~\cite{KM}
except for the rational number in the coefficient of $C_A$ in the
single log term at order $\alpha_s^2$, where we find $-\frac
{55}{108}$ rather than $-\frac {37}{108}$ appearing in Eq. (3.6) of
Ref.~\cite{KM}. This discrepancy is
due to an error in Ref.~\cite{KM} in the evaluation of one of the
diagrams --- see Sec.~\ref{sec:KM} and Appendix~\ref{sec:diagram11}
below.

\subsection{Evolution equation}

From \eq{cal_Soft} above one can be obtain an evolution equation by
taking a full logarithmic derivative with respect to the scale.  We
obtain:
\begin{eqnarray}
\frac{\partial \ln {\cal S}(N\mu/m)}{\partial \ln \mu}+
\frac{\partial \ln {\cal S}(N\mu/m)}{\partial \alpha_s(\mu)}\,
\frac{d\alpha_s(\mu)}{d\ln \mu}&=& \\  \nonumber
\frac{d\ln {\cal S}(N\mu/m)}{\ln \mu}&=& -2{\cal
A}\big(\alpha_s(\mu)\big)\,\ln\frac{N\tilde{\mu}}{m}\,+
\,\Gamma_{\cal S}(\alpha_s(\mu)) \label{ev_eq_S}
\end{eqnarray}
where, as above, $\tilde{\mu}=\mu\,{\rm e}^{\gamma_E}$, and the
anomalous dimension $\Gamma_{\cal S}$ can be related order by order
to the two anomalous dimensions ${\cal D}$ and ${\cal A}$ defined
above:
\begin{equation}
\Gamma_{\cal S}(\alpha_s(\mu))=\frac{C_F}{\beta_0}\int_0^{\infty} du
T(u) \left(\frac{\Lambda^2}{\mu^2}\right)^u\Big[\Gamma(1-2u)B_{\cal
D}(u) \,+\,\frac{1+2u\gamma_E-\Gamma(1-2u)}{u} B_{\cal A}(u) \Big].
\end{equation}
For its expansion we obtain:
\begin{eqnarray}
\Gamma_{\cal
S}(\alpha_s(\mu))&=&C_F\left\{\frac{\alpha_s(\mu)}{\pi}\,+\,
\beta_0\left(\frac{\alpha_s(\mu)}{\pi}\right)^2
\left[d_2^{\MSbar}-\frac{\pi^2}{3}+2\gamma_E-2\gamma_E^2\right]
+\cdots \right\}=\\ \nonumber &&\hspace*{-30pt}= C_F
\left\{\frac{\alpha_s(\mu)}{\pi}+\beta_0
\left(\frac{\alpha_s(\mu)}{\pi}\right)^2
\left[\frac{1}{9}-\frac{\pi^2}{3}+2\gamma_E-2\gamma_E^2+\frac{C_A}{\beta_0}
\left(\frac{9}{4}\zeta_3-\frac{\pi^2}{12}-\frac{11}{18}\right)\right]\right\}.
\end{eqnarray}
Note that \eq{ev_eq_S} is similar but not identical to the equation
derived in Ref.~\cite{KM} for $W[C_S]$ (Eq. 4.5 in Ref. \cite{KM}).
The equation for $W[C_S]$ takes the form
\begin{equation}
\frac{d\ln W[C_S](N\mu/m)}{\ln \mu}= -2{\cal
A}\big(\alpha_s(\mu)\big)\,\ln\frac{N\tilde{\mu}}{m}\,-\Gamma(\alpha_s(\mu)),
\label{ev_eq_W}
\end{equation}
where, according to \eq{def_S_Wilson},
\[\Gamma(\alpha_s(\mu))=-\Gamma_{\cal S}(\alpha_s(\mu))
+\frac{\partial \ln W[C_S](1,\alpha_s(\mu))}{\partial\alpha_s}\,\frac{d\alpha_s}
{d\ln\mu}.
\]
From \eq{MM_vs_KM} we have
\[
\ln W[C_S](1,\alpha_s(\mu))=C_F\frac{\alpha_s(\mu)}{\pi}
\left(-\frac{5}{24}\pi^2+\gamma_E-\gamma_E^2\right)+{\cal
O}(\alpha_s^2),
\]
so the NLO result for $\Gamma(\alpha_s(\mu))$ is:
\begin{eqnarray}
\Gamma(\alpha_s(\mu))&=&-C_F
\left\{\frac{\alpha_s(\mu)}{\pi}+\beta_0
\left(\frac{\alpha_s(\mu)}{\pi}\right)^2
\left[\frac{1}{9}+\frac{\pi^2}{12}+\frac{C_A}{\beta_0}
\left(\frac{9}{4}\zeta_3-\frac{\pi^2}{12}-\frac{11}{18}\right)\right]\right\}
\\ \nonumber &=&-C_F \frac{\alpha_s(\mu)}{\pi}+
C_F\left(\frac{\alpha_s(\mu)}{\pi}\right)^2
\left[\left(\frac{1}{54}+\frac{\pi^2}{72}\right)N_f+\left(\frac{55}{108}
+\frac{\pi^2}{144} -\frac{9}{4} \zeta_3\right)C_A\right].
\end{eqnarray}

\subsection{Relation between distribution and
fragmentation in the Sudakov region\label{sec:relation_to_frag}}

In the following we show that the function controlling the
large-$N$ limit of the perturbative heavy--quark {\em fragmentation
function} is identical to ${\cal S}(N\mu/m)$
of the heavy--quark {\em distribution function}.

According to the discussion in
Sec.~\ref{sec:Wilson_lines} above, the log--enhanced terms in the
distribution function can be computed in the Eikonal approximation,
namely using the Wilson line definition of \eq{def_Wilson}.
Following Sec. 2.3 in Ref.~\cite{KM} we now show that the same
applies in the case of the fragmentation function. We then prove
that in the Sudakov limit the two objects are in fact identical.
This is a new result.

The fragmentation function is defined by~\cite{CS} (see Eq. (5.2) there):
\begin{eqnarray}
\label{D_def} d(x;\mu)&\equiv&
  \frac{x^{1-2\epsilon}}{2\pi} \int_{-\infty}^{\infty} dy_{-}\,
\exp(-i{p^{+}}y^{-}/x)\, \times  \\ \nonumber && \hspace*{-30pt}
\,\frac{1}{4N_c}{\rm Tr}\left\{\sum_{X}\, \gamma_{+} \langle 0 \vert
\, \Psi(0)\,\Phi^{*}_{-y}(\infty,0)\, \,\vert H(p)+X\rangle \langle
H(p)+X\vert\,\Phi^{*}_{y}(y,\infty)\,
\overline{\Psi}(y)\,\vert0\rangle_{\mu}\right\},
\end{eqnarray}
where dimensional regularization in $D=4-2\epsilon$ dimensions is assumed and
$\mu$ is the renormalization scale of the operator. Here
the trace is taken over Dirac and color indices, the sum is
over all hadronic states $X$ that can be produced together with the
observed heavy hadron $H(p)$, $p$ is the momentum of the latter
which is assumed to have no transverse component and $y$ is a
lightlike vector in the ``$-$'' direction. For the Wilson lines we use the
notation
of Eqs. (\ref{path_ordered_exp}) and (\ref{path_ordered_exp_bar}), where
$^*$ stands for complex conjugation;
Ref.~\cite{CS} expresses the same Wilson line in terms of transposed color matrices,
i.e.
\[
\Phi^{*}_{-y}(\infty,0)= \left[\overline{{\bf P}}\,\exp \left(-ig\int_0^\infty dy^{-}
A_a^{+}(y^-)t_a\right)\right]^* =\overline{{\bf P}}\,\exp \left(ig\int_0^\infty dy^{-}
A_a^{+}(y^-)t_a^T\right).
\]
It is straightforward to check that
$\Psi(0)\,\Phi^{*}_{-y}(\infty,0)$ is gauge invariant if the gauge field does not
transform at infinity.
Since $d(x;\mu)$ has support in $x\in [0,1]$ one can define moments as usual,
\begin{equation}
D_N(m;\mu)=\int_0^1dx x^{N-1} d(x;\mu).
\end{equation}
In the perturbative analogue of $d(x;\mu)$, which we denote by $d_{\PT}(x;\mu)$,
$H(p)$ is replaced by an
on-shell heavy quark $h(p)$ with $p^2=m^2$. Contrary to the
non-perturbative definition, in perturbation theory $h(p)$ carries
color, and so does $X$.

Because there are no collinear singularities from the heavy--quark
propagator, Sudakov logarithms are either due to soft gluons, or to
gluons that are collinear with the ``$-$'' direction. Therefore, as
argued in Ref.~\cite{KM}, the Sudakov limit can be safely studied in
the approximation where the dynamical heavy--quark field in each
amplitude is replaced by a free quark multiplied by a Wilson line
along the quark trajectory:
\begin{equation}
\Psi(z)\longrightarrow\psi(z)\,\Phi^*_{p}(z,\infty) \,; \qquad
\bar{\Psi}(z)\longrightarrow\Phi^*_{-p}(\infty,z)\,\bar{\psi}(z)\,,
\label{Quark_to_Wilson_line_frag}
\end{equation}
where the path--ordered exponential is defined as
in~\eq{path_ordered_exp}.
\begin{figure}[ht]
\begin{center}
\epsfig{file=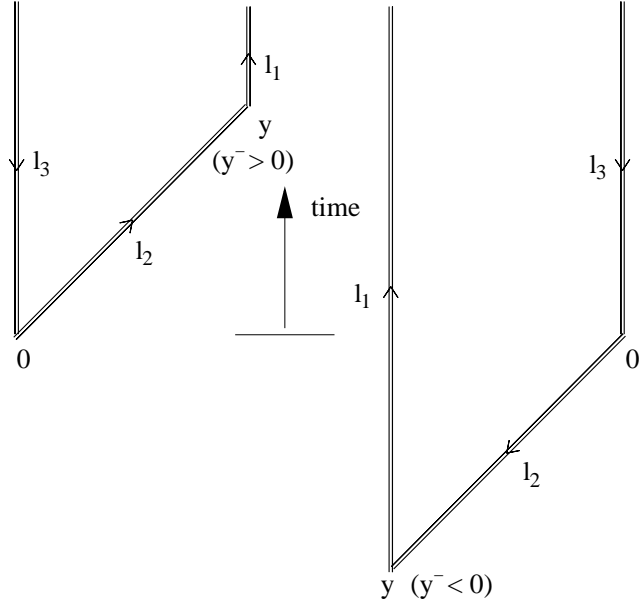,angle=0,width=8.4cm}
\caption{\label{fig:frag}Minkowski space-time picture
of the Wilson--line configuration
in the second line of \eq{D_def_singular}, i.e.
$W^*[C_S](ip\cdot y\mu/m)=W[C_S](-ip\cdot y\mu/m)$,
representing
the perturbative fragmentation function in the infinite--mass limit
(in the rest frame of the produced quark). The two figures describe the situation
when $y^-$ is positive (l.h.s) or negative (r.h.s), where path ordering on the
lightlike segment $l_2$ from $0$ to $y$ corresponds to
time--ordering and anti-time--ordering, respectively;
{\em cf.} Fig. \ref{fig:dist}.
}
\end{center}
\end{figure}

Taking the infinite--mass limit implies that the
detected on-shell quark $h(p)$ is produced from the heavy--quark
field in the operator rather than from some vacuum fluctuation.
Note that when considering the Sudakov limit
such fluctuations are irrelevant {\em even} if the mass is not large, since
they necessarily involve gluon splitting into a
 quark--antiquark pair, which is regular at $x\longrightarrow 1$.
Therefore, in this approximation the free quark field
$\psi$ annihilates $h(p)$ from the external states. Factoring out the
Dirac structure one obtains:
\begin{eqnarray}
&&\frac{1}{4N_c}{\rm Tr}\left\{\sum_{X}\, \gamma_{+} \langle 0 \vert
\, \Psi(0)\,\Phi^{*}_{-y}(\infty,0)\, \,\vert h(p)+X\rangle \langle
h(p)+X\vert\,\Phi^{*}_{y}(y,\infty)\,
\overline{\Psi}(y)\,\vert0\rangle_{\mu}\right\} \,  \longrightarrow
 \nonumber \\ \nonumber
&&\hspace*{40pt}
 p^{+} \,{\rm e}^{ip^{+}y^{-}}
\langle 0 \vert \, \Phi^{*}_{p}(0,\infty) \,\Phi^{*}_{-y}(y,0)\,
\Phi^*_{-y}(\infty,y)\,
\sum_{X}\,\,\vert X\rangle \langle X\vert\,
\Phi_{y}^{*}(y,\infty)\, \Phi_{-p}^*(\infty,y)
\,\vert0\rangle_{\mu}
\\
&&\hspace*{40pt} =
 p^{+} \,{\rm e}^{ip^{+}y^{-}}\bigg[
\langle 0 \vert \, \Phi_{p}(0,\infty) \,\Phi_{-y}(y,0)\,
\Phi_{-p}(\infty,y) \,\vert0\rangle_{\mu}\bigg]^{*},
\end{eqnarray}
where we relied on completeness of the set of states
$\vert X\rangle$ (which close the color trace) and on the properties of the Wilson lines in
\eq{Wilson_properties}. We therefore find that the $x\longrightarrow
1$ singular terms in the fragmentation function are summarized by:
\begin{eqnarray}
\label{D_def_singular} \hspace*{-20pt}\left.d_{\PT}(x;\mu)\right\vert_{x\longrightarrow
1}&\sim&
   \int_{-\infty}^{\infty} \frac{p^{+} dy^{-}}{2\pi}\,
{\rm e}^{-i{p^{+}}y^{-}(1-x)}\, \langle 0 \vert \, \Phi_{p}(0,\infty)
\,\Phi_{-y}(y,0)\, \Phi_{-p}(\infty,y)
\,\vert0\rangle^{\dagger}_{\mu}\nonumber \\
&=&\int_{-\infty}^{\infty}
\frac{p^{+} dy^{-}}{2\pi}\, {\rm e}^{-i{p^{+}}y^{-}(1-x)}\, \langle 0 \vert
\, \Phi_{p}(y,\infty) \,\Phi_{y}(0,y)\,
\Phi_{-p}(\infty,0) \,\vert0\rangle_{\mu} \nonumber\\
&=&\int_{-\infty}^{\infty}
\frac{-q^{+} dy^{-}}{2\pi}\, {\rm e}^{i{q^{+}}y^{-}(1-x)}\, \langle 0 \vert
\, \Phi_{-q}(y,\infty) \,\Phi_{y}(0,y)\,
\Phi_{q}(\infty,0) \,\vert0\rangle_{\mu} \nonumber\\
&=& \int_{-\infty}^{\infty}\frac{-q^{+}dy^{-}}{2\pi}\, {\rm
   e}^{iq^{+}y^{-}(1-x)}\,W[C_S](iq\cdot y  \mu /m,\alpha_s(\mu))\nonumber\\
   &=& \int_{-\infty}^{\infty}\frac{p^{+}dy^{-}}{2\pi}\, {\rm
   e}^{-ip^{+}y^{-}(1-x)}\,W[C_S](-ip\cdot y  \mu /m,\alpha_s(\mu)),
\end{eqnarray}
where in the third line we defined $q\equiv -p$ and in the fourth we
identified the
matrix element as $W[C_S]$ that was defined in the context of the {\em distribution
 function}, Eq. (\ref{def_Wilson}) above. Finally, in the last line we returned
 to the original variable $p$ finding that the function
 $W[C_S]$ is evaluated at $-ip\cdot y  \mu /m$.
 Since $d_{\PT}(x;\mu)$ is real, sign inversion of the argument of $W[C_S]$
is equivalent to its complex conjugation ({\em cf.} \eq{complex_conjugation}) so
we get
\begin{equation}
\label{d_PT_like_f}
\left.d_{\PT}(x;\mu)\right\vert_{x\longrightarrow
1}\sim
\int_{-\infty}^{\infty}\frac{p^{+}dy^{-}}{2\pi}\, {\rm
   e}^{ip^{+}y^{-}(1-x)}\,W[C_S](ip\cdot y  \mu /m,\alpha_s(\mu)),
\end{equation}
the same expression as for $f_{\PT}(x;\mu)$ in \eq{definition_singular}.
Therefore, we find that as far as the ${x\longrightarrow 1}$ terms are
 concerned, $d_{\PT}(x;\mu)$ is identical to $f_{\PT}(x;\mu)$.
Having the relation in the second line \eq{D_def_singular} above,
Ref.~\cite{KM} has defined
\begin{equation}
\label{W_C_T}
W[C_T](ip\cdot y\mu/m,\alpha_s(\mu))\equiv \langle 0 \vert
\, \Phi_{p}(y,\infty) \,\Phi_{y}(0,y)\,
\Phi_{-p}(\infty,0) \,\vert0\rangle_{\mu}.
\end{equation} Here we find that $W[C_T]$ is related to
$W[C_S]$ by complex conjugation:
\begin{equation}
W[C_T](ip\cdot y\mu/m,\alpha_s(\mu)) = W[C_S]^{*}(ip\cdot y\mu/m,\alpha_s(\mu)).
\label{WCT_WCS}
\end{equation}

Finally, converting \eq{d_PT_like_f} to moment space
$ip^{+}y^{-}\longrightarrow N$ (see
Eq. (9) in Ref.~\cite{CG}):
\begin{eqnarray}
\label{F_W_relation_D} D_N^{\PT}(m;\mu) =H_W^D(m;\mu) \,\times\, W[C_S](N  \mu
/m,\alpha_s(\mu))+{\cal O}(1/N),
\end{eqnarray}
so one obtains the {\em same Sudakov exponent as in the distribution
function to any logarithmic accuracy}. The $N$-independent terms are
summarized by $H_W^D(m;\mu)$. At ${\cal O}(\alpha_s)$ $H_W^D(m;\mu)$ is
equal to $H_W^F(m;\mu)$  of \eq{HW}, but this may not persist at
higher orders.

\subsection{Comments on the calculation of the Sudakov exponent to NNLO\label{sec:KM}}

In this section we explain how the two--loop coefficient
$d_2^{\MSbar}$ of \eq{d_2} was obtained. Having established the
all--order equality between the Sudakov exponents of the heavy quark
distribution and fragmentation functions as well as the relation
between the definition based on a
dynamical heavy quark with a finite on-shell mass and
the Wilson--line definition, there are several way to proceed. We follow two:
\begin{itemize}
\item{} Perform a two-loop calculation using Wilson lines, as done in
Ref.~\cite{KM} by Korchemsky and Marchesini.
\item{} Extract the non-Abelian\footnote{The $C_F N_f$ term was
known already
in Ref.~\cite{CG}, and was confirmed by~Ref.\cite{MM}.}
$N\longrightarrow \infty$ singular terms from a recent result for
the fragmentation function by Melnikov and Mitov~\cite{MM}, which
uses dynamical heavy quarks with a finite mass.
\end{itemize}

Beginning with the latter, the two-loop calculation of
Ref.~\cite{MM} conveniently suites our purpose, as it relies on a
{\em process--independent definition} of the perturbative fragmentation
function in dimensional regularization~\cite{CC}, where process dependent
power corrections in the
hard scale (the scale at which the heavy quark is produced) are
avoided by taking the {\em quasi--collinear
limit}~\cite{Catani:2000ef,CC}. In this limit the gluon transverse
momentum and the quark mass are taken {\em small} while the ratio
between them, which depends on the quark longitudinal momentum
fraction, is fixed. Ref.~\cite{CC} established this definition and applied it
at ${\cal O}(\alpha_s)$, conforming previous result \cite{MN} which was obtained
from heavy--quark production cross-section in $e^+e^-$ annihilation.
Ref.~\cite{CC} also presented results for Sudakov resummation to NLL
accuracy. Ref.~\cite{CG} extended the process--independent
calculation of the fragmentation function to all orders in the
large--$\beta_0$ limit, which, in particular, fixes the Abelian part of
$d_2^{\MSbar}$, see \eq{dn_an_relation} and Table \ref{Table_coeff} above.

Ref.~\cite{MM} gives a general two-loop result for $d_{\PT}(x;\mu)$ in
momentum fraction space. It also summarizes in moment space the terms which are
non-vanishing at large $N$ in Eq.~(65). We note\footnote{I wish to
thank Matteo Cacciari and Kiril Melnikov for related discussions.}
that the coupling in this paper is renormalized assuming
$N_f+1$ dynamical massless quarks, where the additional flavor
corresponds to the heavy quark. This gives rise to $C_FT_R
\alpha_s^2$ terms which are not accompanied by $N_f$. Converting to
our definition of the coupling, with $N_f$ light quarks, these
contributions drop out. The remaining terms in Eq. (65) of Ref.~\cite{MM}
match the general expression of \eq{soft_mom_def_my} with Eqs.
(\ref{Soft}) and (\ref{H_}) above {\em provided} that $d_2^{\MSbar}$ is
given by \eq{d_2}.

Here we performed\footnote{As mentioned following \eq{MM_vs_KM}, the
result quoted in Ref.~\cite{KM} differs from my own in the non-Abelian
coefficient of the single log
term, the one that determines $d_2$. Historically, it was
the discrepancy with $d_2$ I
extracted from Ref.~\cite{MM} which convinced me to repeat the
two-loop calculation of Ref.~\cite{KM}. I~wish to thank
Gregory Korchemsky for his encouragement.} a two--loop calculation
of $W[C_S]$ of \eq{def_Wilson}, along the lines of
Ref.~\cite{KM}.
We recall that $W[C_S]$ is a path--ordered exponential with two
antiparallel rays in the timelike directions $n_{\mu}$ and $-n_{\mu}$,
which are connected by a finite lightlike segment $y_{-}$.
The timelike direction $n_{\mu}$ is determines by the heavy--quark momentum: $p_{\mu}=m n_{\mu}$,
{\em but} in contrast with the original definitions, Eqs. (\ref{definition}) and
(\ref{D_def}), and with the calculation of Ref.~\cite{MM} discussed above,
the heavy quark is no more a dynamic field. It is replaced by a Wilson line that
represents interaction with soft gluons only (collinear singularities do arise
though from the lightlike segment).
As shown in the previous sections $W[C_S]$ captures the log-enhanced terms
in the heavy quark distribution and fragmentation functions to all orders,
i.e. to any logarithmic accuracy. Working in $D=4-2\epsilon$ dimensions in
configuration space with the Feynman rules as in Appendix A of
Ref.~\cite{KM} the standard ${\overline {\rm MS}}$ scale $\mu$ is
introduced by making the following replacements:
\begin{eqnarray}
\label{MSbar_scale_def}
\begin{array}{lll}
&& n\cdot y\longrightarrow n\cdot y \,\mu\,, \\
&& \frac{g^2}{4\pi^2} \longrightarrow
\frac{\alpha_s^{\MSbar}(\mu)}{\pi} \, \left(\frac{{\rm e}^{
\gamma_E}}{4\pi}\right)^{\epsilon}
\left(1-\frac{\alpha_s^{\MSbar}(\mu)}{\pi}\frac{\beta_0}{\epsilon}+{\cal
O}(\alpha_s^2)\right).
\end{array}
\end{eqnarray}

Our final result for the renormalized $\ln W[C_S]$ to ${\cal O}(\alpha_s^2)$ is
summarized by \eq{MM_vs_KM} above. Given that the description of the
calculation in Ref.~\cite{KM} is clear and detailed, we shall not
repeat it here. Instead, we give a full account of the
calculation of one diagram, diagram~11, the one by which we differ
from Ref.~\cite{KM}. This is done in Appendix~\ref{sec:diagram11}.

\section{Conclusions}

In this paper we studied perturbative aspects of
the heavy--quark distribution function, which has a central role in
precision phenomenology of inclusive $B$--decay spectra, primarily in
$\bar{B}\longrightarrow X_s\gamma$ and semileptonic decays.
Our results include all--order resummation of running--coupling effects,
in \eq{N_space_result}, as well as determination of the Sudakov exponent in
\eq{Soft_Catani} to two-loop order.
The result for $d_2^{\MSbar}$ is \eq{d_2} is now established in two
entirely different calculation procedures: the one performed here
following Ref.~\cite{KM} using a Wilson--line
operator in configuration space and the one of
Ref.~\cite{MM} for the fragmentation function, using the quasi--collinear
limit~\cite{CC,Catani:2000ef} in momentum space.
With the jet-function anomalous dimension already being known to this
order~(see e.g.~\cite{GR}), the Sudakov exponent in inclusive decay spectra
is now determined to the NNLL accuracy. As usual, this should be matched
by the computation of $N$-independent terms at ${\cal O}(\alpha_s^2)$, which
are not yet available.

In addition to presenting the result for Sudakov resummation in the conventional way
(\eq{Soft_Catani}) that suites
fixed--logarithmic--accuracy calculations,
following previous work on DGE~\cite{BDK,DGE_thrust,Gardi:2001di,CG} we formulated
the resummation in~\eq{Soft} as a scheme--invariant Borel sum,
where power-like separation between perturbative and non-perturbative contributions
can be implemented by taking a Principal Value prescription.
This approach is particularly advantageous for inclusive $B$--decay spectra where non--perturbative
corrections are substantial. The application of this approach to phenomenology
is already under way~\cite{AG}.

The quark distribution function is defined here
 assuming a finite on-shell quark mass, while much of the literature
 on inclusive decays is based on
defining it in the $m\longrightarrow \infty$ limit.
The evolution properties of these objects
are different.
Starting in Sec.~\ref{sec:Wilson_lines} with the QCD Sudakov resummation formula
for the case of a finite on-shell mass
we derived a strictly factorized form that is consistent with the $m\longrightarrow \infty$ limit.
In the former the Sudakov factor is naturally defined with
factorization--scale independent normalization. This is realized in \eq{Soft_Catani}
and in \eq{Soft}.
This is not the case in the strictly--factorized formula of~\eq{cal_Soft}, where
the first moment strongly depends on the scale.
Strict factorization (\eq{strict_fact}) {\em implies} that both the Sudakov factor~${\cal S}$
and the hard factor~${\cal H}$ acquire {\em double logarithmic} dependence on the
scale to any order in perturbation theory (see e.g. \eq{HW}).
Therefore, in this case {\em both} need to be resummed.

The most interesting finding of our present investigation is the similarity of the
distribution and fragmentation functions. First we found that
in the large--$\beta_0$ limit these functions are identical. Then, we showed that in
the Sudakov limit they are represented by the {\em same} Wilson--line operator
--- see \eq{D_def_singular} ---
so the {\em Sudakov exponent in the two cases is identical to all orders.}
We emphasize that the diagrammatic realization of this relation is non
trivial: upon calculating separately virtual corrections to the fragmentation process one
encounters additional Coulomb--phase contributions in individual diagrams, that are absent in the
distribution case. Our result implies that these contributions cancel out in the sum of
all diagrams so they make no effect on the Sudakov exponent.

In spite of these strong relations the distribution and fragmentation functions,
defined in a process--independent way in dimensional regularization, are not equal.
Their DGLAP evolution away from the large--$x$ limit starts
differing already at two-loop order. Additional differences between these functions
 appear when considering
heavy quark--antiquark pairs that were neglected here; these are important in the
case of fragmentation for low masses and away from the large-$x$ limit.
In spite of the
similarity in the renormalon structure it is hard to imagine that there is any
relation between power corrections in the two cases. Recall that
the distribution function is defined with a single hadron in the initial state and
a completely inclusive final state, making it a forward hadronic matrix element. On the other hand
in the case of fragmentation both the detected hadron and the jet are
in the final state and they interact by exchanging soft gluons throughout
the hadronization~process.

\section*{Acknowledgements}

I wish to thank Gregory Korchemsky for very helpful
discussions and for checking and verifying my result for diagram 11
of Ref.~\cite{KM}, and Carola Berger for collaboration during some
of the difficult stages of this project.
I would like to thank the KITP at UC Santa Barbara for hospitality
while some of this work was done.
The work is supported by a Marie Curie individual fellowship, contract number
HPMF-CT-2002-02112.

\appendix

\section{Properties of Wilson lines\label{Wilson_lines}}

We define the Wilson--line operator by
\begin{equation}
\Phi_p(z_1,z_2) \equiv {\bf P} \exp\left(ig\int_{z_1}^{z_2}dz_{\mu}
A^{\mu}(z)\right). \label{path_ordered_exp}
\end{equation}
with $A^{\mu}(z)=A^{\mu}_{a}(z) t_{a}$ where $t_{a}$ are ${\rm SU}(N_c)$
generators ($t_a^{\dagger}=t_a$) in the fundamental representation and
${\bf P} \exp$ indicates that matrices and fields are path--ordered.
The notation $\Phi_p(z_1,z_2)$ assumes\footnote{There is some
redundancy in the notation as $z_2-z_1$ is parallel to $p$, but it is, nevertheless,
convenient.} that the direction $p$
is from $z_1$ to $z_2$. Anti-path--ordering is denoted by:
\begin{equation}
\Phi_{-p}(z_2,z_1) \equiv {\overline {\bf P}} \exp\left(-ig\int_{z_1}^{z_2}dz_{\mu}
A^{\mu}(z)\right). \label{path_ordered_exp_bar}
\end{equation}
The following properties of $\Phi_p(z_1,z_2)$ are useful:
\begin{eqnarray}
\label{Wilson_properties}
\begin{array}{lll}
&{\rm causality}  &\Phi_p(z_2,z_3)\Phi_p(z_1,z_2)=\Phi_p(z_1,z_3)
 \qquad {\rm where}\,\, z_2\,\, {\rm is\,\, between}\,\, z_1\,\, {\rm and}\,\,
  z_3\\
&{\rm hermiticity}\qquad &\Phi_p^{\dagger}(z_1,z_2)=\Phi_{-p}(z_2,z_1)\\
&{\rm unitarity}  &\Phi_p^{\dagger}(z_1,z_2) \Phi_p(z_1,z_2)=1.
\end{array}
\end{eqnarray}

\section{One-loop integrals with a Borel--modified propagator}

Let us compute the quark distribution on an on-shell heavy quark
with a single dressed gluon based the definition of \eq{definition}.
First we note that the numerator is identical to the one computed in
Sec. 3.1 of Ref.~\cite{CG} with the replacement of ${\cal
M}_0\bar{{\cal M}}_0$ by~$\nsl$, where $n$ is in the ``$-$''
direction. Upon using a Borel--modified gluon propagator
as well as dimensional regularization we get:
\begin{eqnarray}
\label{using_def}
   \left.f_{\PT}(x;\mu)\right|_{{\cal O}(\alpha_s)}
   &=&\int_{-\infty}^{\infty}\frac{dy^{-}}{4\pi}\, {\rm   e}^{-ixp^{+}y^{-}}\,
   \left< h(p) \right\vert\bar{\Psi}(y) \Phi_{y} (0,y) \gamma_{+} \Psi(0)
   \left\vert h(p)
    \right>_{\mu} \\ \nonumber
 &=&
\frac{-i4 C_Fg^2}{\pi}\,p^{+} \int_{-\infty}^{\infty}\frac{dy^{-}}{4\pi}\, {\rm
e}^{-ixp^{+}y^{-}}
\int\frac{d^Dk}{(2\pi)^D}
{\rm
e}^{iy^{-}(p^{+}+k^{+})}
\\ \nonumber  &&\hspace*{50pt}\times
\Bigg[\frac{{\frac{k^{+}}{p^{+}}}+2
+2{\frac{p^{+}}{k^{+}}}}{(-k^2)^{1+u}((p+k)^2-m^2)}
-\frac{(2m^2+k^2)(1+{\frac{k^{+}}{p^{+}})}}{(-k^2)^{1+u}((p+k)^2-m^2)^2}\Bigg],
\end{eqnarray}
where the $\frac{k^{+}}{p^{+}}$ part in the first term and the
entire second term originate in the Feynman gauge part of the gluon
propagator and the rest is specific to the Axial gauge
(\ref{lightcone_gauge}).

To perform the momentum integration we use the fact that under the
$y^{-}$ integral, a factor of $\frac{-k^{+}}{p^{+}}$  in the
numerator becomes $(1-x)$ while the inverse factor becomes
$1/(1-x)$. We therefore have to deal with just one type of integral:
\begin{equation}
\label{I} I(m^2,x;a,b,D)\equiv
p^{+}\int_{-\infty}^{\infty}\frac{dy^{-}}{2\pi}\, {\rm
e}^{-ixp^{+}y^{-}}\,\int\frac{d^Dk}{(2\pi)^D}{\rm
e}^{iy^{-}(p^{+}+k^{+})} \frac{1}{(-k^2)^a((p+k)^2-m^2)^b}.
\end{equation}
Using Feynman parametrization and shifting the integration momentum
$q=k+p(1-\alpha)$ and changing the order of integration we obtain:
\begin{eqnarray}
\label{I_Fey}
I(m^2,x;a,b,D)&=&p^{+}\frac{(-1)^{-a}\Gamma(a+b)}{\Gamma(b)\Gamma(a)}\,\int_0^1d\alpha
\alpha^{a-1}(1-\alpha)^{b-1}
\int_{-\infty}^{\infty}\frac{dy^{-}}{2\pi}\, {\rm
e}^{-i(x-\alpha)p^{+}y^{-}}\nonumber \\
&&\hspace*{140pt}\times\,\int\frac{d^Dq}{(2\pi)^D}{\rm
e}^{iy^{-}q^{+}} \frac{1}{(q^2-m^2(1-\alpha)^2)^{a+b}}\nonumber \\
\nonumber &=&
\frac{i(-1)^{b}\Gamma(a+b-\frac{D}{2})}{(4\pi)^{\frac{D}{2}}\Gamma(b)\Gamma(a)}\,
\int_0^1d\alpha \delta(\alpha-x)\alpha^{a-1}(1-\alpha)^{b-1}
 (m^2(1-\alpha)^2)^{\frac{D}{2}-a-b}\\
&=& \frac{i(-1)^{b}
\Gamma(a+b-\frac{D}{2})}{(4\pi)^{\frac{D}{2}}\Gamma(b)\Gamma(a)}\,
(m^2)^{\frac{D}{2}-a-b}
\,x^{a-1}(1-x)^{D-1-2a-b},
\end{eqnarray}
where we first performed the momentum integration (observing that
the exponential ${\rm e}^{iy^{-}q^{+}}$  can be replaced by $1$
since the result is a scalar). Then we performed the $y^-$
integration getting a Dirac $\delta(\alpha-x)$. This made the
integration over the Feynman parameter $\alpha$ trivial. We comment
that had we used the Eikonal approximation for the massive
propagator, i.e. $(p+k)^2-m^2\longrightarrow 2pk$ we would have
obtained the same answer (same $\Gamma$ functions) but the
dependence on $x$ would have been modified such that
$x^{a-1}\longrightarrow 1$, not affecting the large--$x$ limit.

\eq{using_def} can now be computed by the appropriate assignments in
\eq{I_Fey}. The result is given by \eq{f_result}.

\section{The splitting function and the cusp anomalous
dimension\label{sec:cusp}}

The non-singlet splitting function has been recently computed to
three loops by Moch, Vermaseren and Vogt~\cite{MVV}. In addition,
its large--$\beta_0$ limit is known to all orders and it is given
by~\cite{Gracey}
\begin{eqnarray}
\label{AP} \gamma(N,a) &=&\frac{C_F}{\beta_0}\sum_{n=0}^{\infty}
 \gamma_n(N) a^{n+1}={\cal A}\,\bigg[ \Psi(N+a)-\Psi(1+a)  \\ \nonumber
&+&\frac{N-1}{2}
\left(\frac{a^2+2a-1}{1+a}\,\frac{1}{N+a}-\frac{(1+a)^2}{2+a}\frac{1}{N+1+a}
\right) \bigg]\,+\,{\cal O}(1/\beta_0^2)
\end{eqnarray}
where ${\cal A}$ is the large-$\beta_0$ limit of the cusp anomalous
dimension~\cite{Korchemsky:1988si,KR87,KR92,KM,Beneke:1995pq} discussed below, and
$a\equiv a(\mu)=\beta_0\alpha_s(\mu)/\pi$ is the large-$\beta_0$
coupling in $\overline{\rm MS}$.

The perturbative expansion of the cusp anomalous dimension (the
anomalous dimension of an operator made of two Wilson lines with a
cusp), which is also the large--$N$ limit of the quark--quark
splitting function, is given by
\begin{eqnarray}
\label{A_cusp} {\cal A}\big(\alpha_s(\mu)\big)&=&
\frac{C_F}{\beta_0}\,\int_0^{\infty} du \,T(u)
\,\left(\frac{\Lambda^2}{\mu^2}\right)^u\, B_{\cal A}(u) \nonumber \\
&=&
\frac{C_F}{\beta_0}\bigg[\left(\frac{\beta_0\alpha_s^{\MSbar}(\mu)}{\pi}\right)
+a_2^{\MSbar}
\left(\frac{\beta_0\alpha_s^{\MSbar}(\mu)}{\pi}\right)^2
+a_3^{\MSbar}
\left(\frac{\beta_0\alpha_s^{\MSbar}(\mu)}{\pi}\right)^3
+\cdots\bigg],
\end{eqnarray}
where the coefficient $a_3^{\MSbar}$ is known from the recent
calculation of the splitting function~\cite{MVV}. Explicitly,
\begin{eqnarray}
\label{a23_MSbar}
 a_2^{\MSbar}&=& {\displaystyle \frac {5}{3}}  +
{\displaystyle {\frac{\mathit{C_A}}{\beta_{0}} \left({\displaystyle
\frac {1}{3}}  - {\displaystyle \frac {\pi ^{2}}{12}} \right)\,}},
\\ \nonumber
 a_3^{\MSbar}&=&  - {\displaystyle \frac {1}{3}}  + \frac{1}{\beta_0}\left[
{\displaystyle  {\left({\displaystyle \frac {55}{16}}  - 3\,\zeta_3
 \right)\,\mathit{C_F} + \left({\displaystyle \frac {253}{72}}  -
{\displaystyle \frac {5\,\pi ^{2}}{18}}  + {\displaystyle \frac {
7}{2}} \,\zeta_3\right)\,\mathit{C_A}}}  \right]\nonumber \\
\nonumber \mbox{}&&\hspace*{30pt} +  \frac{1}{{\beta_0}^2}
\left[\left( - \frac {605}{192}  +  \frac {11}{4} \,\zeta_3\right)
\,\mathit{C_A}\,\mathit{C_F}\,+\, \left( - {\displaystyle \frac
{7}{18}}  - {\displaystyle \frac {\pi ^{2}}{18}} - {\displaystyle
\frac {11}{4}} \,\zeta_3 +  \frac {11\,\pi ^{4}}{720}
\right)\,\mathit{C_A}^{2} \right].
\end{eqnarray}

In the large--$\beta_0$ limit~\cite{Gracey,Beneke:1995pq}:
\begin{equation}
{\cal A}\big(\alpha_s(\mu)\big) =  \frac{C_F}{\beta_0}\,\frac{\sin
\pi a}{\pi}\,\,\frac{\Gamma(4+2a)}{6\Gamma(2+a)^2}\,+ \,{\cal
O}(1/\beta_0^2). \label{G_large_Nf}
\end{equation}

The Borel representation of ${\cal A}\big(\alpha_s(\mu)\big)$ in the
full theory can be written in an expanded form as in
\eq{BA_expanded}, where $c_n$ represent the terms that are
subleading in $\beta_0$. Upon comparing the latter with the second
line in \eq{A_cusp} and using \eq{tHooft_coupling} we get:
\begin{eqnarray}
c_2&=&a_2^{\MSbar}-\frac53\nonumber \\
c_3&=&a_3^{\MSbar}+\frac 13+\delta_2^{\MSbar}-\delta a_2^{\MSbar},
\end{eqnarray}
where the term involving $\delta_2=\beta_2^{\MSbar}/\beta_0^3$ is
due to converting from $\overline{\rm MS}$  to the 't Hooft scheme
(see Eq. (27) in \cite{GR}). Explicitly this gives:
\begin{eqnarray}
c_2&=&\frac{C_A}{\beta_0}\left(\frac{1}{3}-\frac{\pi^2}{12}\right)
\nonumber \\
c_3&=& \frac{1}{\beta_0}
\left[\left(\frac{649}{288}-\frac{5}{18}\pi^2
+\frac{7}{2}\zeta\left(3\right)\right)C_A
+\left(\frac{23}{8}-3\zeta_3\right)C_F\right] \nonumber \\
&+&\frac{1}{\beta_0^2}
\left[\left(\frac{251}{288}+\frac{7}{144}\pi^2
-\frac{11}{4}\zeta_3+\frac{11}{720}\pi^4\right) C_A^2
+\left(-\frac{235}{96}+\frac{11}{4}\zeta_3+\frac{\pi^2}{16}\right)C_FC_A
-\frac{3}{32}C_F^2\right]\nonumber \\
&+&\frac{1}{\beta_0^3}
\left[\left(-\frac{301}{512}-\frac{7}{192}\pi^2\right)
C_A^3+\left(-\frac{11}{64}
-\frac{11}{192}\pi^2\right)C_FC_A^2+\frac{11}{128}C_F^2C_A\right].
\end{eqnarray}
The presence of a ${\cal O}(1/{\beta_0^3})$ term is, of course, a
special feature of our specific Borel representation (or of the 't
Hooft scheme).

\section{Calculation of diagram 11 in Ref.~\cite{KM}\label{sec:diagram11}}

\begin{figure}[th]
\begin{center}
\epsfig{file=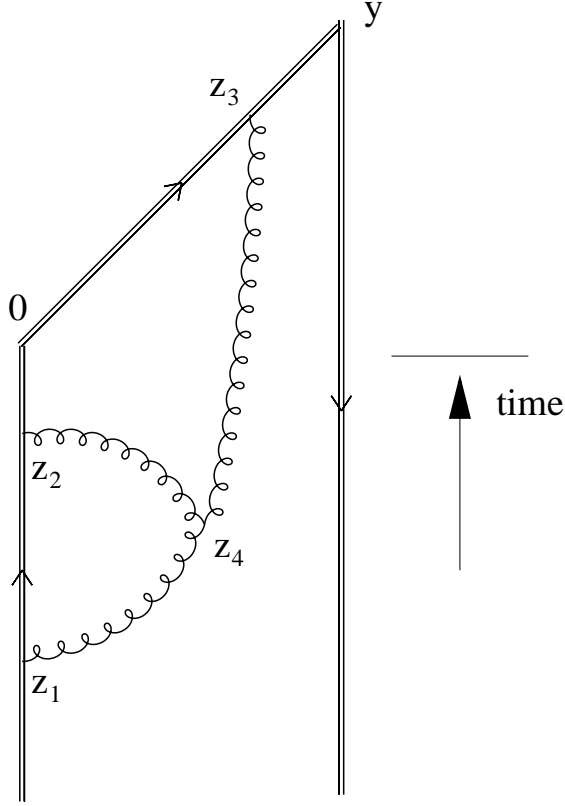,angle=0,width=7.4cm}
\caption{\label{fig:KM11}One of the non-Abelian two-loop diagrams contributing to
$\ln W[C_S]$ of \eq{def_Wilson}. It corresponds to diagram 11 in Fig.~6 of Ref.~\cite{KM}
(published version). }
\end{center}
\end{figure}

Let us focus here on the calculation of the non-Abelian two-loop diagram
shown in Fig.~\ref{fig:KM11} in the Feynman gauge.
The calculation is done in configuration space using dimensional
regularization in $D$
space-time dimensions. The Feynman rules are given in Appendix A in
Ref.~\cite{KM}. This diagram involves a triple gluon vertex, where two
gluons attach at different points $z_1$ and $z_2$ along the time
like Wilson line representing the incoming heavy quark (with
momentum $p_{\mu}=m n_{\mu}$; $n^2=1$) and the third gluon attaches
to the lightlike line going along the ``$-$'' direction at the
point $z_3$.

We parametrize the points along the Wilson lines as in
Ref.~\cite{KM}, $z_1=\tau_1 n$, $z_2=\tau_2 n$ and $z_3=\tau_3 y$,
and denote the triple gluon vertex by $z_4$, obtaining:
\begin{eqnarray}
W_{11}=\frac{-i}{2} g^4 C_AC_F
\int_{-\infty}^0d\tau_1\int_{\tau_1}^0d\tau_2\int_0^1d\tau_3\,
\,\Big(y - (n\cdot y)n\Big)  \cdot
\left(\frac{d}{dz_1}-\frac{d}{dz_2}\right) {\cal J}(z_i)
\label{W_11_rules}
\end{eqnarray}
with
\begin{equation}
 {\cal J}(z_i) \equiv \int\, d^Dz_4 \,\prod_{i=1}^3 \,\,{\bf D}(z_i-z_4)
\end{equation}
where the Feynman gauge propagator ${\bf D}(z)$ is:
\begin{eqnarray}
{\bf D}(z)=\int\frac{d^Dk}{(2\pi)^D}\,{\rm e}^{-ik\cdot z}
\frac{i}{k^2+i\varepsilon}
=\frac{\Gamma(D/2-1)}{4\pi^{D/2}}\bigg[-z^2+i\varepsilon \bigg]^{1-D/2}.
\label{full_propagator}
\end{eqnarray}
The calculation of ${\cal J}(z_i)$ using Feynman parametrization is
straightforward and yields:
\begin{equation}
 {\cal J}(z_i) = \frac{-i\Gamma(D-3)}{4^3\pi^D}\int_0^1 dt
 \int_0^{1-t} ds \left[st(1-s-t)\right]^{D/2-2}
 \left(-\Delta\right)^{3-D}
\end{equation}
with
\begin{eqnarray}
\Delta &\equiv& t(1-t) z_1^2 + s(1-s) z_2^2 +(s+t)(1-s-t)z_3^2\nonumber \\
&&-2st z_1\cdot z_2-2s(1-s-t) z_2\cdot z_3-2t (1-s-t) z_1\cdot z_3\nonumber \\
&=& \tau_1^2t(1-t)-2st \tau_1 \tau_2+\tau_2^2s(1-s)
 -2n\cdot y (1-s-t) \tau_3\left[s \tau_2+t \tau_1 \right].
\end{eqnarray}
Taking the derivative in \eq{W_11_rules} and scaling $\tau_{1,2}$ by
$-2n\cdot y$ one obtains:
\begin{eqnarray*}
&&W_{11}=\frac12 g^4C_AC_F\,\frac{\Gamma(D-2)(2in\cdot
y)^{8-2D}}{2^7\pi^D} \int_0^1\!ds\int_0^{1-s}\!\!dt (1-s-t)^{D/2-1}
(st)^{D/2-2}(s-t)
 \nonumber \\&&
\!\!\int_0^{\infty}\!\!d\tau_1\int_0^{\tau_1}d\tau_2\!\!\int_0^1\!\!d\tau_3
\tau_3\bigg[ \tau_1^2t(1-t)-2st \tau_1 \tau_2+\tau_2^2s(1-s)
+ (1-s-t) \tau_3 \left(s \tau_2+t \tau_1\right)\bigg]^{2-D}.
\end{eqnarray*}
By first changing the integration variable $\tau_2$ into $\sigma$,
where $\tau_2=\sigma \tau_1$, and then the integration variable
$\tau_1$ into $\omega$, where
\[
\tau_1=\omega \frac{(1-s-t)\tau_3(\sigma
s+t)}{\sigma^2s(1-s)-2st\sigma+t(1-t)},
\]
one finds that the $\tau_3$ integration is trivial, giving a factor
of $1/(8-2D)$, while the $\omega$-integration yields:
\[
\int_0^{\infty} d\omega \omega^{3-D}
(1+\omega)^{2-D}=-\frac{4^D\Gamma(D-5/2)
\Gamma(3-d)}{128\sqrt{\pi}}.
\]
The result can be expressed as:
\begin{eqnarray}
\label{W_11_non_ren} W_{11}=\frac12 g^4 C_A C_F g^4 \frac{(2in\cdot
y)^{8-2D}}{4^2\pi^D} \,\,I_{11},
\end{eqnarray}
where $I_{11}$ matches the notation of Ref~\cite{KM}. We have:
$I_{11} = K_{11} \times J_{11}$ with
\begin{eqnarray}
J_{11}&=& \int_0^1d\sigma\int_0^1 ds\int_{0}^{1-s} dt\,
(1-s-t)^{5-\frac32D}(st)^{\frac{D}{2}-2}(s-t)\times\nonumber \\
&&\hspace*{80pt}(\sigma s+t)^{6-2D}
\left(\sigma^2s(1-s)-2st\sigma+t(1-t)\right)^{D-4},
\end{eqnarray}
and
\begin{equation}
\label{K_11} K_{11}=-
\frac{4^{D-5}\Gamma(D-2)\Gamma(D-5/2)\Gamma(3-D)}
{(8-2D)\sqrt{\pi}}.
\end{equation}

Proceeding with the evaluation of $J_{11}$ we change the integration
variable $t$ into $z$, where $t=(1-s)z/(1+z)$, and then, after
changing the order of integration, $\sigma$ into $x$ where
$x=(1+z)(1-\sigma)$ and $s$ into $y$ where $y=s/z$, getting:
\begin{equation}
J_{11}=\int_0^{\infty}\!\!dz (1+z)^{D-5}\int_0^{1+z}\!\!\!\!dx
\int_0^{1/z}\!\!\!\!dy y^{D/2-2} (y(1+2z)-1)
\Big(1+y(1-x)\Big)^{6-2D} \Big(1+y(1-x)^2\Big)^{D-4}.
\end{equation}
Since the dependence of the integrand on $z$ is simple we perform
this integration first. The price is having several terms as the
$z$-integration extends between $\max\{0,x-1\}$ and $1/y$.
Fortunately, most of the terms cancel out by symmetry and we obtain:
\begin{eqnarray}
\label{J_11_intr} J_{11}&=& \frac{-2}{D-3}\int_0^{\infty}
dy\int_1^{(1+y)/y}\!\!\! dx
y^{D/2-1}x^{D-3}\Big(1+y(1-x)\Big)^{6-2D} \Big(1+y(1-x)^2\Big)^{D-4}
\\ \nonumber &+& \frac{1}{D-4} \int_0^{\infty}
dy\int_1^{(1+y)/y}\!\!\! dx (1+y)y^{D/2-2}x^{D-4}
\Big(1+y(1-x)\Big)^{6-2D} \Big(1+y(1-x)^2\Big)^{D-4}.
\end{eqnarray}
By changing the integration variable $x$ into $z$ where
$z=1+(1-z)/y$ and then $y$ into $w$ where $y=(1-z)/w$, the two
integrals can be computed exactly, yielding:
\begin{eqnarray}
\label{J_11_result}
J_{11}&=& \frac{-2}{D-3}\, \frac{\pi}{\sin (\pi D/2)}\,
\frac{\Gamma(6-3D/2)}{\Gamma(3-D/2)\Gamma(6-D)}\\
\nonumber &+& \frac{1}{D-4}\,\frac{1}{(2D-7)}\,\frac{\pi}{\sin (\pi
D/2)}\left[
{}_3F_2\left(\begin{array}{lll}4-D,7-2D,2-D/2\\
8-2D,6-3D/2\end{array},1\right)\right.\\\nonumber &&\hspace*{170pt}
\left. - {}_3F_2\left(\begin{array}{lll}
4-D,7-2D,1-D/2\\
8-2D,7-3D/2\end{array},1\right)\right],
\end{eqnarray}
where the first line is the result of the first integral in
\eq{J_11_intr} while the remaining terms, containing hypergeometric
functions, correspond to the second.

Combining Eqs. (\ref{J_11_result}) and (\ref{K_11}) we obtain the
final result for $I_{11}$, which can be readily expanded in
$\epsilon$, where $D=4-2\epsilon$. The result is:
\begin{eqnarray}
I_{11}&=&\frac{1}{192}\epsilon^{-4}
+\left(-\frac{1}{96}+\frac{1}{96}\gamma_{E}\right)\epsilon^{-3}
+\left(-\frac{1}{48}+\frac{13}{1152}\pi^2+\frac{1}{96}\gamma_{E}^2
-\frac{1}{48}\gamma_{E}\right)
\epsilon^{-2} \\
\nonumber&&\hspace*{10pt}+\left(\frac{19}{288}\zeta_3+\frac{13}{576}
\pi^2\gamma_{E}
+\frac{1}{144}\gamma_{E}^3-\frac{1}{24}-\frac{1}{48}
\gamma_{E}^2-\frac{1}{24}\gamma_{E}-\frac{13}{576}
\pi^2\right)\epsilon^{-1}+{\cal O}(1).
\end{eqnarray}
Finally, in terms of the renormalized coupling, \eq{W_11_non_ren}
takes the form:
\begin{eqnarray}
W_{11}&=&\frac12 C_A C_F
\left(\frac{\alpha_s^{\MSbar}(\mu)}{\pi}\right)^2 \,
(in\cdot y\mu{\rm e}^{\frac12 \gamma_E})^{4\epsilon} \,\,I_{11}
\nonumber \\&=&
\frac12 C_A C_F \left(\frac{\alpha_s^{\MSbar}(\mu)}{\pi}\right)^2
\, {\rm e}^{4 L \epsilon} \,\,{\rm e}^{-2\gamma_E\epsilon} I_{11},
\end{eqnarray}
where in the first line we introduced the ${\overline{\rm MS}}$
factorization and renormalization scale according
to~\eq{MSbar_scale_def} and in the second we defined $L\equiv \ln
ip\cdot y\tilde{\mu}/m = \ln N \tilde{\mu}/m$ with
$\tilde{\mu}\equiv {\rm e}^{\gamma_E} \mu$, as in \eq{MM_vs_KM},
knowing that the $\epsilon$-expansion of ${\rm
e}^{-2\gamma_E\epsilon} I_{11}$ is free of $\gamma_E$ terms.
Performing this expansion, and subtracting the
$\epsilon\longrightarrow 0$ singular terms, we finally get:
\begin{eqnarray}
W_{11}^{\rm ren.}&=&\frac12 C_A C_F
\left(\frac{\alpha_s^{\MSbar}(\mu)}{\pi}\right)^2
\left[\frac{1}{18}L^4-\frac{1}{9}L^3+\left(\frac{13}{144}\pi^2
-\frac{1}{6}\right)L^2\right. \\
&&\nonumber\hspace*{130pt}+\left.\left(\frac{19}{72}
\zeta_3-\frac{1}{6}-\frac{13}{144}\pi^2\right)L+{\cal
O}(1) \right],
\end{eqnarray}
which differs from Eq. (3.6) in Ref.~\cite{KM} by the single-log
term only.

\end{document}